\documentclass[a4paper,12pt]{article}

\usepackage{amsmath, amsthm, amsfonts}
\usepackage[lmargin=.75in,rmargin=.75in,tmargin=1.in,bmargin=1in]{geometry}
\usepackage{acronym}
\usepackage{graphicx}
\usepackage{caption}
\usepackage{subcaption}
\usepackage[usenames, dvipsnames]{color}
\usepackage[square, sort, numbers]{natbib}
\usepackage{float} 
\acrodef{HIV}{Human Immunodeficiency Virus}
\acrodef{HIV-1}{Human Immunodeficiency Virus type 1}
\acrodef{AIDS}{Acquired Immunodeficiency Syndrome}
\acrodef{ELISA} {Enzyme-Linked Immunosorbent Assay}
\acrodef{ART} {antiretroviral therapy}
\acrodef{AI} {Avidity Index}
\acrodef{PA} {Population-Averaged}
\acrodef{GEE} {Generalized Estimating Equations}
\acrodef{MSE} {Mean Squared Error}
\usepackage{csquotes}
\usepackage{lscape}
\usepackage{comment}
\usepackage{rotating}
\usepackage{tikz}
\usetikzlibrary{arrows, decorations.markings, matrix}
\usetikzlibrary{positioning}
\usetikzlibrary{fit,shapes.misc}
\usetikzlibrary{arrows}
\usetikzlibrary{shapes}
\usepackage{mathtools}
\usepackage{physics}

\usepackage{tikz}
\usepackage{siunitx}
\usetikzlibrary{decorations.pathreplacing,,positioning}
\tikzset{mybox/.style = {rectangle, text width=4cm,minimum height=2cm,align=center},
mybox2/.style = {rectangle, text width=6cm,minimum height=2cm,align=center}
}
\usepackage{tikz}
\usetikzlibrary{arrows, decorations.markings, matrix}
\usetikzlibrary{bayesnet}
\usepackage{pdfpages}

\newcommand{\bmath}[1]{\mbox{ \boldmath $\!#1\!$ \unboldmath}}
\let\newcommand=\providecommand

\title{Quantifying the recency of HIV infection using multiple longitudinal biomarkers}

\author{Loumpiana Koulai, Anne Presanis, Gary Murphy, Barbara Suligoi \\ and Daniela De Angelis } 

\begin{document}
\maketitle
Knowledge of the time at which an HIV-infected individual seroconverts, when the immune system starts responding to HIV infection, plays a vital role in the design and implementation of interventions to reduce the impact of the HIV epidemic. A number of biomarkers have been developed to distinguish between recent and long-term HIV infection, based on the antibody response to HIV. To quantify the recency of infection at an individual level, we propose characterising the growth of such biomarkers from observations from a panel of individuals with known seroconversion time, using Bayesian mixed effect models. We combine this knowledge of the growth patterns with observations from a newly diagnosed individual, to estimate the probability seroconversion occurred in the X months prior to diagnosis. We explore, through a simulation study, the characteristics of different biomarkers that affect our ability to estimate recency, such as the growth rate. In particular, we find that predictive ability is improved by using joint models of two biomarkers, accounting for their correlation, rather than univariate models of single biomarkers.

\section{Introduction} \label{sec:intro}
Following infection with the Human Immunodeficiency Virus (HIV), the
immune system responds by producing anti-HIV antibodies of different
types at different stages from infection \citep{chen2002}, culminating
in what is known as seroconversion, {\it i.e} the time at which
antibodies are detectable in blood serum. CD4 counts and viral load traditionally have been used as prognostic biomarkers of HIV progression \citep{lange1992,mellors1997} but have been less successful for estimating time since infection, due to their non-monotonic behaviour and the difficulty of observing individuals at the early stages of infection . In recent years, focussing instead on the antibody response, a number
of serological assays, able to detect different aspects of this diverse response, have been developed with the goal of distinguishing recent from long-standing infections. Typically, for a specified biomarker,  a threshold is chosen and HIV positive individuals with a measured optical density (OD) below the threshold are classified as recently infected (see \cite{suligoi2011,parekh2011,laeyendecker2012,keating2012,kassanjee2014,kassanjee2016bench} and references therein). This classification has  been used to estimate HIV incidence at population level  \citep{karon2008,levu2012}.
At an individual level, however, this dichotomization does not allow a
clear quantification of the recency of infection. The type of statement that can be made on recency can only be on average,  based on knowledge of the mean time taken to cross the chosen threshold
from seroconversion.

\begin{figure}[H]
\centering
\includegraphics[scale=0.5]{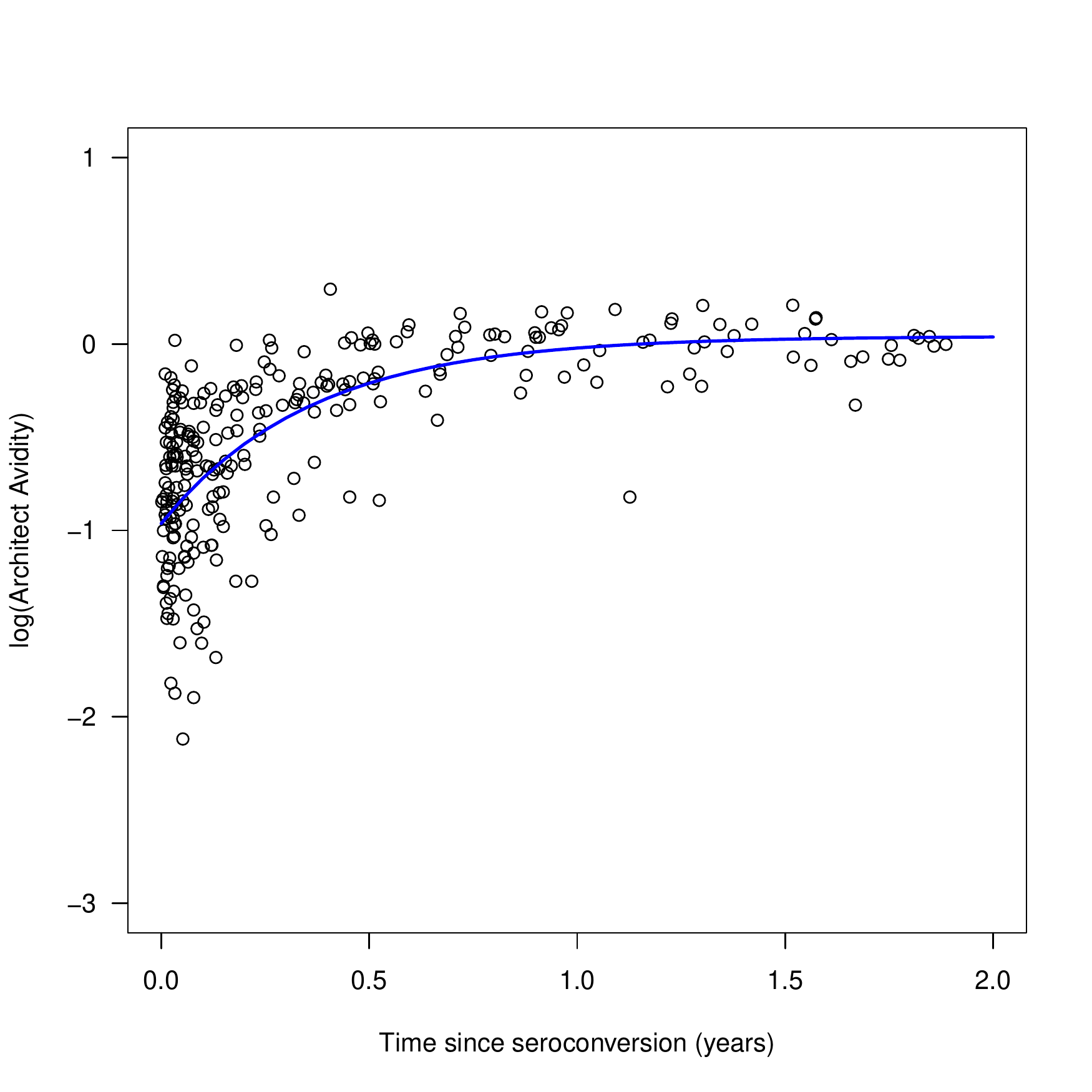}
\caption{Architect Avidity (on logarithmic scale) with the blue line representing the mean growth pattern.}
\label{fig:gof_italian}
\end{figure}

The key question of interest is whether it is possible to make probabilistic statements at individual-level about the recency of HIV infection. Can biomarker measurements for a newly diagnosed individual, combined with knowledge of the natural evolution of the biomarker, be used to infer the probability, $P_X$, that an individual has seroconverted in the $X$ months prior to diagnosis? Antibody-response biomarkers increase monotonically and approach a plateau over time. An example is the Architect Avidity \citep{suligoi2011}, whose growth pattern is shown in Figure~\ref{fig:gof_italian}. These data are routinely collected from HIV-positive patients attending clinics in Italy. These patients have known (or well-estimated) seroconversion times, and at each clinic visit following diagnosis, one or more biomarkers are measured. For such a panel of individuals, the growth pattern of each biomarker observed can be estimated. Figure~\ref{fig:gof_italian} shows each individual's OD values of Architect Avidity against time since seroconversion, with the estimated mean growth curve in blue. Given such growth curves and observations on a newly diagnosed individual, how well can we estimate the seroconversion probability $P_X$?

$P_X$ can be derived from the distribution of an individual's seroconversion time. Up to now, little attention has been paid to developing methods for estimating individual seroconversion times. Traditionally, the midpoint between the last negative and the first positive HIV test date has been adopted as an estimate \citep{alcabes1993}. A number of authors have considered the use of markers of immune response to improve the estimation of seroconversion time. \cite{munoz1989,munoz1992} fit a Weibull model to the known seroconversion times of seroconverters, using CD4 counts as a covariate. The fitted model is then used to impute the unknown seroconversion times for seroprevalent individuals for whom CD4 counts are available. \cite{dubin1994} develop a Bayesian model for estimating the conditional distribution of time since seroconversion given CD4 counts at the time of the first positive HIV test. 
More recently, \cite{sommen2011} model the evolution of two non-linearly evolving biomarkers by using the same functional form for each
biomarker, deriving parameter estimates through a maximum likelihood approach. Resulting estimates are then
used via Bayes' rule to estimate the distribution of infection time for each individual in their sample. Similarly,
\cite{konikoff2015} uses a Bayesian bivariate non-linear mixed-effects model,
with the same functional form for each antibody-response biomarker, to
estimate the average time spent in the recent infection state. More recently, \cite{borremans2016} models separately the level
of a measured biomarker and presence/absence of recency, assuming
that they are independent. These two sources of information are
combined into one conditional probability that the time of
infection is recent. However, in reality, levels or presence of different
biomarkers may be correlated and this correlation should be taken into account in the estimation process.

Our aim is to explore the feasibility of using a limited number of serial measurements of one or more biomarkers to quantify the recency of HIV infection for any newly diagnosed patient. Univariate linear and non-linear mixed-effect models to describe the growth patterns of antibody response and viral load biomarkers are given in Section~\ref{sec:methods1}. Joint non-linear mixed-effects models of bivariate biomarkers are given in Section~\ref{sec:methods2}. We evaluate the performance of single and multiple intrinsically correlated biomarkers in estimating the probability $P_X$ of having seroconverted in the $X$ months prior to HIV diagnosis through a simulation study (Section~\ref{sec:simulation}). Biomarkers with different growth patterns are investigated to evaluate the impact of particular characteristics, such as the growth rate, on the accuracy of the estimation. 
Results are reported in Section~\ref{sec:results} and we end with a discussion in Section~\ref{sec:discussion}.

\section{Biomarker models} \label{sec:methods}
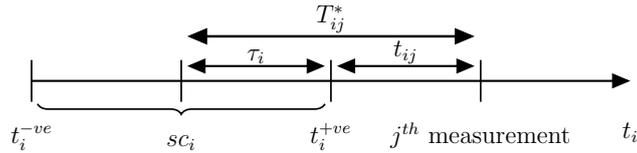
\begin{figure}
\centerline{
\resizebox{0.5\textwidth}{!}{%
\begin{tikzpicture}
\draw[->,line width=1] (0,0)node[pos=0,below=0.55]{ $t_i^{-ve}$} --
node[pos=0.5,below=0.55]{$t_i^{+ve}$}
node[pos=0.375,above=3.85pt]{ $\tau_i$}
node[pos=0.75,below=0.55]{$j^{th}$ measurement}
node[pos=0.625,above=3.85pt]{ $t_{ij}$}
node[pos=1,below=0.55]{$t_i$}
node[pos=0.5,above=18.85pt]{ $T_{ij}^{*}$}
node[pos=0.25,below=0.55pt,mybox2]{$sc_i$}
(10,0);
\draw[thick]  (0,-10pt)--(0,10pt) (2.5,-10pt)--(2.5,10pt)  (5,-10pt)--(5,10pt)  (7.5,-10pt)--(7.5,10pt);
\draw[<->, line width=1](4.9,0.25)--(2.6,0.25);
\draw[<->, line width=1](7.4,0.25)--(5.1,0.25);
\draw[<->, line width=1.2](7.4,0.75)--(2.6,0.75);
\draw[decorate,thick,decoration={brace,amplitude=5pt,raise=10pt},] 
(4.9,-1pt)--(0.1,-1pt);
\end{tikzpicture}}
}
\caption{Time-span of the sequence of measurements for each individual $i$.}
\label{fig:time_scale}
\end{figure}

Let $y_{ij}^k$ denote the observed measurement of the random variable $Y^k$ representing the $k^{th}$ biomarker, for the $i^{th}$ individual at the $j^{th}$ observation time, $t_{ij}$, where $i=1, \dots, n$, $j=1, \dots, n_i$, $k=1, \dots, K$. Assume that the available data for $n$ individuals (see Figure~\ref{fig:time_scale}) also include the dates of the last negative and the first positive HIV test, $t_i^{-ve}$ and $t_i^{+ve}$ respectively. The interval $[t_i^{-ve},t_i^{+ve}]$ is the interval within which individual $i$ has seroconverted, with length $sc_i=t_i^{+ve}-t_i^{-ve}$. Note that $\tau_i$ is the time from seroconversion to $t_i^{+ve}$ and $T_{ij}^*=\tau_i+t_{ij}$ is the time from seroconversion to the $j^{th}$ measurement.

\subsection{Single outcome models} \label{sec:methods1}
Suppose that a single outcome $k$ is measured on each individual $i$
at each time point $t_{ij}$ from the first positive HIV test date. The
observed longitudinal trajectories of biomarker $k$ can be modelled as
\begin{equation}
y_{ij}^k=g(t_{ij}+\tau_i,\bmath{\beta}_i^k)+\epsilon_{ij}^k  
\label{eq:generic_univ}
\end{equation}
where $\epsilon_{ij}^k \sim N(0,\sigma_{\epsilon^k}^2)$ represent
normally distributed measurement errors. Function $g(\cdot)$
represents the true underlying values of biomarker $k$ and depends on
the time since seroconversion $(t_{ij}+\tau_i)$ and random effects
$\bmath{\beta}_i^k$, that are normally distributed with mean 0 and
variance-covariance matrix $\Sigma_{\beta^k}$. Different functional forms of $g(\cdot)$ can be used to capture the underlying evolution of a biomarker of interest. In what follows markers of antibody-response and viral presence will be considered.

\subsubsection{Antibody response}\label{sec:methods1.1}
Antibody response may evolve linearly over time since seroconversion \citep{keating2016}. Such evolution can be represented by a linear mixed-effects model with random intercept $\beta_{1i}^k$ and random slope $\beta_{2i}^k$ \citep{diggle2013book,verbeke2000book,fitzmaurice2008book}:
\begin{equation}
g(t_{ij}+\tau_i,\bmath{\beta}_i^k)=\beta_{1i}^{k} + \beta_{2i}^{k}(t_{ij}+\tau_i)
\label{eq:antibody_response_linear}
\end{equation} 
The intercept represents the value of the biomarker at seroconversion
and the slope the growth rate.

Alternatively, and more commonly, antibody response follows a non-linear trajectory \citep{suligoi2002,suligoi2011,laeyendecker2012,keating2012}. The three-parameter non-linear function used by \citet{sweeting2010} could be adopted to describe the growth of monotonically increasing biomarkers:
\begin{equation}
g(t_{ij}+\tau_i,\bmath{\beta}_i^k)=\beta_{1i}^{k} + (\beta_{2i}^{k} - \beta_{1i}^{k}) * exp\big(-exp(\beta_{3i}^{k})(t_{ij}+\tau_i)\big)
\label{eq:antibody_response}
\end{equation}
This function has intercept $\beta_{2i}^k$ and approaches an asymptote $\beta_{1i}^{k}$ over a period of time. The parameter $\beta_{3i}^{k}$ is the logarithm of the rate constant, representing the growth rate for each individual $i$.

\subsubsection{Viral presence}\label{sec:methods1.2}
Viral presence is thought to be exponentially decreasing and approaching a plateau within a short period after seroconversion \citep{mellors1996,mellors1997,ledergerber1999,saag1996}, as the immune response starts controlling the infection. A two-parameter exponential decay function could be used to model such trajectories:
\begin{equation}\label{eq:viral_load}
g(t_{ij}+\tau_i,\bmath{\beta}_i^k)=\beta_{1i}^{k}\Big(1+exp\big(-\beta_{2i}^{k}(t_{ij}+\tau_i)\big)\Big)
\end{equation}
This non-linear function has decay rate $\beta_{2i}^{k}$ and plateau $\beta_{1i}^{k}$.

\subsection{Bivariate outcome models}\label{sec:methods2}
Suppose now that two outcomes are measured on each individual $i$ over
time from the first positive HIV test date. The response vector for an individual $i$ at time $t_{ij}$ is $(y_{ij}^1,y_{ij}^2)^T$ with $\bmath{y}_i^k=(y_{i1}^k,y_{i2}^k, \dots, y_{in_{i}}^k)^T$ being the sequence of measurements for each biomarker $k$ and $\bmath{t}_i=(t_{i1},t_{i2}, \dots, t_{in_i})^T$ being the sequence of measurement times. A bivariate joint model for the response outcomes is:
\begin{equation} \label{eq:joint_model}
\begin{pmatrix}
\bmath{y}_{i}^1\\
\bmath{y}_{i}^2
\end{pmatrix}
=  \begin{pmatrix}
\bmath{g_{1}}(\tau_i,\bmath{t}_i,\bmath{\beta}_i^1)\\
\bmath{g_{2}}(\tau_i,\bmath{t}_i,\bmath{\beta}_i^2)
\end{pmatrix}
+ \begin{pmatrix}
\bmath{\epsilon}_{i}^1\\
\bmath{\epsilon}_{i}^2
\end{pmatrix}
\end{equation}
where the $\bmath{\epsilon}_{i}^1, \bmath{\epsilon}_{i}^2$ are the within-subject measurement errors of the first and second biomarker respectively and
\begin{equation*} 
 \bmath{g_{k}}(\tau_i,\bmath{t}_i,\bmath{\beta}_i^k)=
 \begin{pmatrix}
 g_{k}(t_{i1}+\tau_i,\bmath{\beta}_i^k)\\
g_{k}(t_{i2}+\tau_i,\bmath{\beta}_i^k)\\
\vdots\\
g_{k}(t_{in_i}+\tau_i,\bmath{\beta}_i^k)\\
 \end{pmatrix}, \qquad k = 1,2.
\end{equation*}
The measurement errors are assumed to be independent and normally distributed with mean $\bmath{0}$ and variance-covariance matrix $\Sigma_{\epsilon}=\begin{pmatrix}
\sigma_{\epsilon^1}^2\bmath{I}_{n_i^1} & \bmath{0}\\
\bmath{0}&\sigma_{\epsilon^2}^2\bmath{I}_{n_i^2}
\end{pmatrix}$
where $\bmath{I}_{n_i}$ denotes the $n_i\times n_i$ identity matrix.
The random effects $\bmath{\beta}_i^1$, $\bmath{\beta}_i^2$ follow the joint multivariate normal distribution with mean vector $\mu_{\bmath{\beta}}$ and variance covariance matrix $\Sigma_{\bmath{\beta}}=\begin{pmatrix}
\Sigma_{\beta}^1&\Sigma_{\beta}^{12}\\
\Sigma_{\beta}^{21}&\Sigma_{\beta}^2
\end{pmatrix}$,
which is partitioned into four sub-matrices: (a) $\Sigma_{\beta}^1$
includes variances and covariances of the random effects for biomarker $1$, (b) $\Sigma_{\beta}^2$ includes variances and covariances
of the random effects for biomarker $2$, (c)
$\Sigma_{\beta}^{12}=\Sigma_{\beta}^{21}$ includes covariances between
random effects of each biomarker, allowing for correlation between the two
biomarkers.

We first consider a bivariate outcome consisting of a linearly and a non-linearly evolving antibody-response biomarker, for example two different avidity assays \citep{chawla2007,suligoi2011}. Additionally, a bivariate outcome of a non-linearly evolving antibody-response biomarker and viral load is examined.

\section{Simulation study}{\label{sec:simulation}}

Assume data from a panel of 100 HIV-infected \enquote{in-sample}
individuals with known seroconversion times $(\tau_i)$ are
available. For each individual, measurements of biomarkers of antibody
response and viral load are taken at HIV diagnosis and regularly
thereafter. The information provided by the \enquote{in-sample}
individuals is used to model the dynamics of biomarkers of interest. A
new \enquote{out-of-sample} individual, with \textit{unknown}
seroconversion time in the seroconversion interval $[t_i^{-ve},t_i^{+ve}]$, is diagnosed in a healthcare facility. For this
new individual, a number of biomarkers are measured at HIV diagnosis
and every few weeks thereafter. 

\subsection{Generating simulated datasets}{\label{sec:simulation_generating}}
We simulate 100 datasets, each consisting of 100 \enquote{in-sample} and 5
\enquote{out-of-sample} individuals. All \enquote{in-sample}
individuals are assumed to be observed every three months from the
first positive HIV test date to two years thereafter, resulting in
nine observed values of univariate and bivariate outcomes. For each
new individual we generate single and bivariate measures at HIV
diagnosis, two weeks and one month afterwards. We use smaller time
intervals between consecutive measurements for new individuals to
investigate the feasibility of recency quantification within a reasonably short period
after HIV diagnosis. The length of the seroconversion interval is
assumed to be one year for both the \enquote{in-sample} and \enquote{out-of-sample} individuals. The seroconversion time for the \enquote{in-sample} individuals is
generated uniformly from the seroconversion interval, $\tau_i \sim U(0,1), i \in 1,\ldots,100$. The seroconversion time for the \enquote{out-of-sample} individuals is set to be five days (0.014 years), three months (0.250 years), six months (0.500 years), nine months (0.750 years) and 360 days (0.986 years) respectively before the first positive HIV test date.

\begin{figure}
\centerline{
\includegraphics[width=9cm,height=7.5cm]{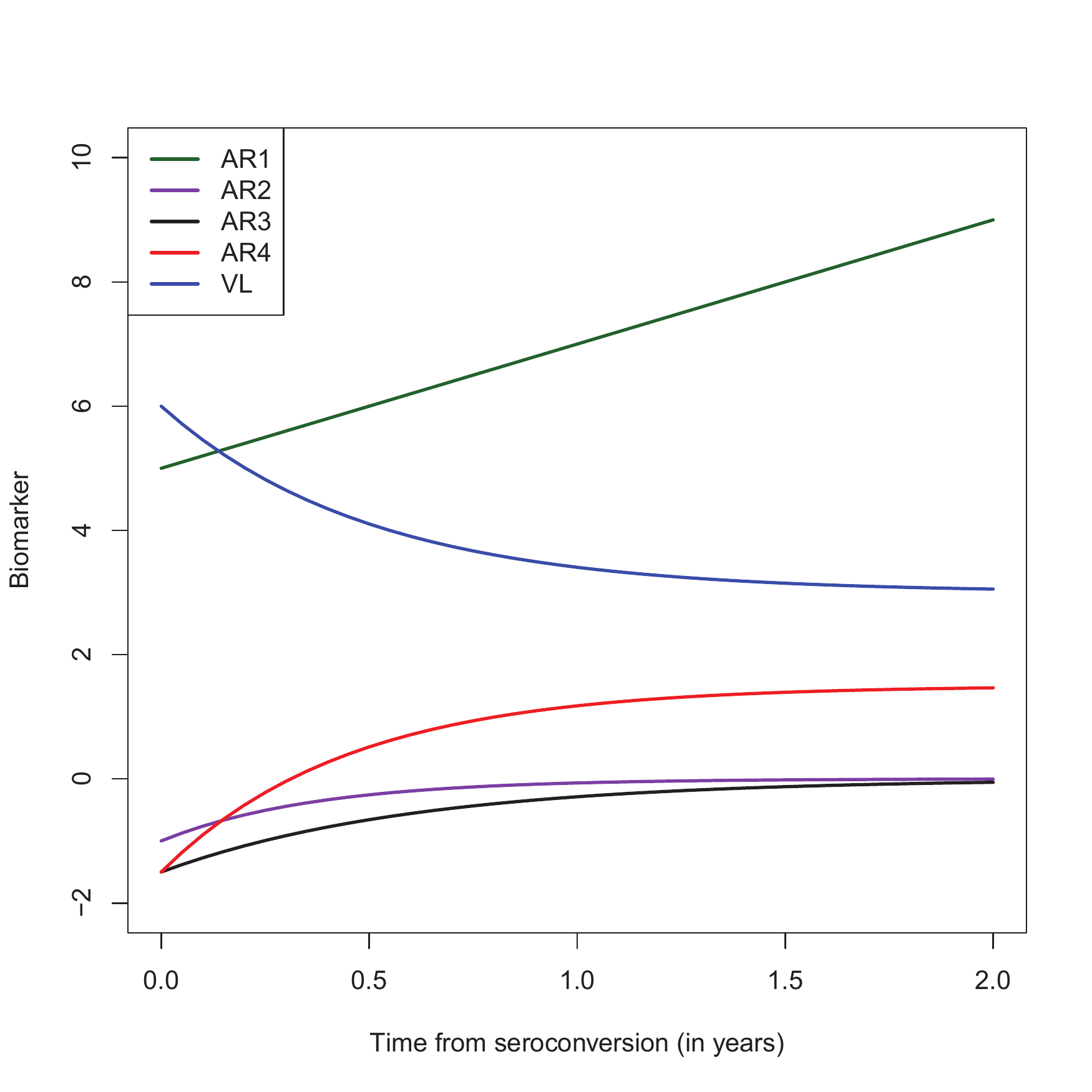}}
 \caption{Mean dynamics of univariate outcomes. AR1 is linearly-evolving, whereas AR2, AR3 and AR4 represent non-linearly evolving antibody-response biomarkers.}
 \label{fig:underlying_trajectories}  
          \end{figure}

Univariate and bivariate realizations of antibody response and viral
load are generated by using equations~(\ref{eq:generic_univ}) and
(\ref{eq:joint_model}) with values of growth parameters as in
Table~\ref{tab:gne_values}. In this \enquote{\textbf{realistic
    scenario}}, the mean and variance of the random effects for
univariate outcomes are chosen so as to resemble the log-transformed
trajectories of existing biomarkers of antibody response, such as the
Avidity Index, LAg Avidity and viral load
\citep{alcabes1993,suligoi2002,suligoi2011,laeyendecker2012,keating2012,keating2016}. We
generate one linearly (AR1) and three non-linearly (AR2, AR3, AR4)
evolving biomarkers of antibody response with their mean trajectories
shown in Figure~\ref{fig:underlying_trajectories}. The asymptote of
each non-linearly evolving antibody-response biomarker is a fixed effect for all individuals. Biomarkers AR2, AR3, and AR4 differ in terms of their growth rates, asymptotes and intercepts. In particular, biomarker AR4 is steeper compared to biomarkers AR2 and AR3 and has a higher asymptote.

\begin{figure}
\centerline{
  \includegraphics[width=9cm,height=7.5cm]{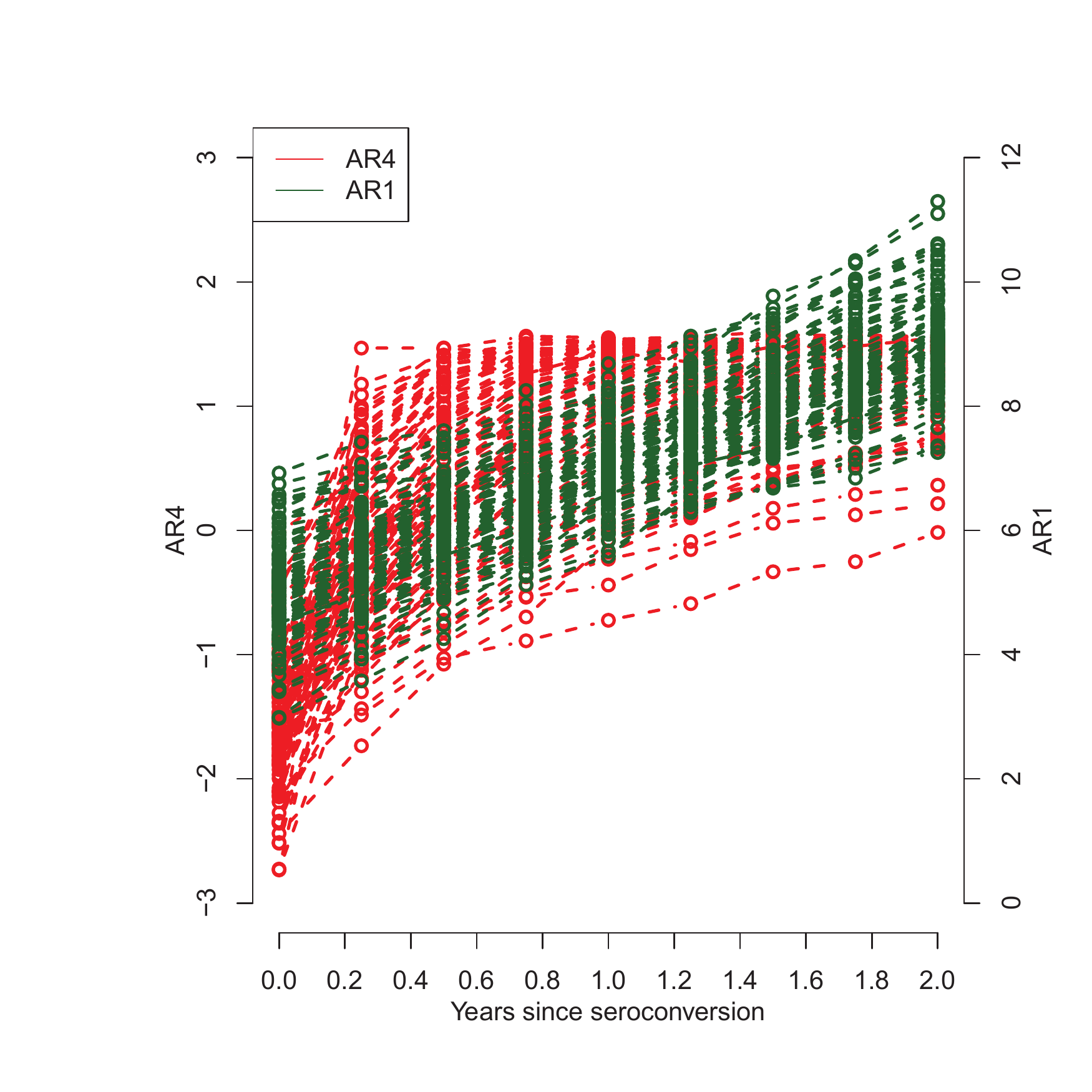}}
\hfill
\centerline{
  \includegraphics[width=9cm,height=7.5cm]{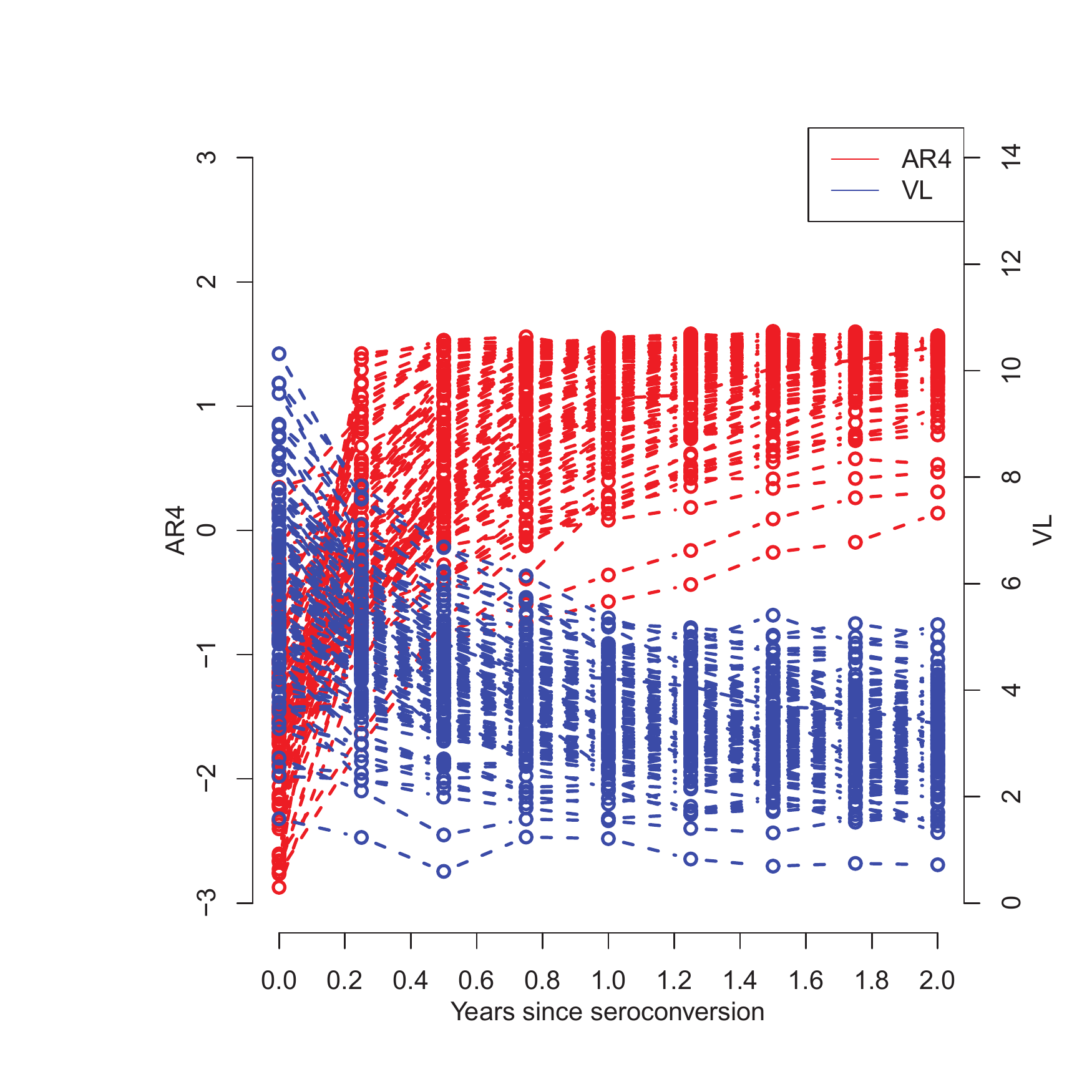}}
\caption{Subject-specific trajectories of bivariate outcomes as generated for one simulated dataset.}
\label{fig:real_trajectories}  
\end{figure}

Bivariate outcomes of two antibody-response biomarkers are
assumed to have
  positively correlated intercepts ($\rho =
  0.1$) and growth rates ($\rho = 0.5$, see
Table~\ref{tab:gne_values}). High initial viral load may trigger rapid growth
  of antibodies; conversely, high initial antibody response may induce
a rapid decline in viral load. We therefore assume that viral load
intercepts and antibody growth rates are positively correlated ($\rho
= 0.3$); as are antibody intercepts and viral load declines ($\rho = 0.3$). Subject-specific trajectories of bivariate outcomes, generated using equation~(\ref{eq:joint_model}) and parameter values presented in Table~\ref{tab:gne_values}, are shown in Figure~\ref{fig:real_trajectories}.

The growth model parameters may affect the ability of
antibody-response biomarkers and viral load to quantify the recency of
HIV infection. For linearly evolving biomarkers, where the slope
defines the change in biomarker values over time, a steep slope
results in mean values differing at consecutive time points. Thus, the
mean at each time point may be strongly related to a particular
seroconversion time. For non-linearly evolving biomarkers, given a
fixed asymptote, the intercept along with the growth rate define the
time that the asymptote will be approached. In particular, a rapidly
evolving biomarker will approach the asymptote within a short period
after seroconversion. After the asymptote is approached, such a biomarker will no longer be
discriminative of recency. In contrast, a slowly evolving biomarker
will approach the asymptote a long time after seroconversion. However,
the mean of such a biomarker may be very similar
at different time points, with values that are challenging
  to relate to particular seroconversion times.

\begin{table}
\caption{Parameter values for generating univariate and bivariate outcomes under the realistic scenario.}
\label{tab:gne_values}
\centering
\scalebox{0.7}{
\begin{tabular}{crrr}
  \hline
Model&Mean & Variance-Covariance & Measurement error\\
& & matrix of random effects & \\
\hline
Antibody Response&&&\\
AR1 & $\mu_{\beta^{AR1}}=(5,2)$&$\Sigma_{\beta^{AR1}}=\begin{pmatrix} 
0.5000 & -0.1900 \\ 
 -0.1900 & 0.2000  
\end{pmatrix} $&$\sigma_{\epsilon^{AR1}}=0.1000$\\
AR2 & $\mu_{\beta^{AR2}}=(0,-1,1)$&$\Sigma_{\beta^{AR2}}=\begin{pmatrix} 
0.0000 & 0.0000 & 0.0000 \\ 
0.0000 & 0.2000 & -0.0850 \\ 
0.0000 & -0.0850 & 0.4000  
\end{pmatrix} $&$\sigma_{\epsilon^{AR2}}=0.0500$\\
AR3 & $\mu_{\beta^{AR3}}=(0,-1.5,0.5)$&$\Sigma_{\beta^{AR3}}=\Sigma_{\beta^{AR2}}$&$\sigma_{\epsilon^{AR3}}=0.0500$\\
AR4 & $\mu_{\beta^{AR4}}=(1.5,-1.5,0.8)$&$\Sigma_{\beta^{AR4}}=\begin{pmatrix} 
0.0000 & 0.0000 & 0.0000 \\ 
0.0000 & 0.4000 & -0.1470 \\ 
0.0000 & -0.1470 & 0.6000  
\end{pmatrix} $&$\sigma_{\epsilon^{AR4}}=0.0500$\\

Joint model AR1 \& AR4 & $\mu_{\bmath{\beta}}=(1.5,-1.5,0.8,5,2)$&$\Sigma_{\bmath{\beta}}=\begin{pmatrix} 
 0.0000 & 0.0000 & 0.0000 & 0.0000 & 0.0000 \\ 
0.0000 & 0.4000 & -0.1470 & 0.0450 & -0.0280 \\ 
0.0000 & -0.1470 & 0.6000 & -0.0550 & 0.1730 \\ 
0.0000 & 0.0450 & -0.0550 & 0.5000 & -0.1900 \\ 
0.0000 & -0.0280 & 0.1730 & -0.1900 & 0.2000 \\
\end{pmatrix} $&$\Sigma_{\bmath{\epsilon}}=\begin{pmatrix} 
0.0025 & 0  \\ 
0 & 0.0100
\end{pmatrix} $ \\
   \hline
Viral load&&&\\
VL  & $\mu_{\beta^{VL}}=(3,2)$&$\Sigma_{\beta^{VL}}=\begin{pmatrix} 
1.0000 & 0.3536  \\ 
0.3536 & 0.5000
\end{pmatrix} $&$\sigma_{\epsilon^{VL}}=0.2000$\\
Joint model AR4 \& VL & $\mu_{\bmath{\beta}}=(1.5,-1.5,0.8,3,2)$&$\Sigma_{\bmath{\beta}}=\begin{pmatrix} 
0.0000&0.0000&0.0000&0.0000&0.0000\\
0.0000&0.4000&-0.1470&0.0630&0.1340\\
0.0000&-0.1470&0.6000&0.2320&0.0550\\
0.0000&0.0630&0.2320&1.0000&0.3536\\
0.0000&0.1340&0.0550&0.3536&0.5000\\
\end{pmatrix} $&$\Sigma_{\bmath{\epsilon}}=\begin{pmatrix} 
0.0025 & 0  \\ 
0 & 0.0400
\end{pmatrix}$\\
   \hline   
\end{tabular}}
\end{table}

\subsection{Analysing simulated datasets}{\label{sec:simulation_analysis}}
The analysis is conducted in a Bayesian framework, using a Markov chain Monte Carlo (MCMC) algorithm as implemented in OpenBUGS 3.2.3 \citep{lunn2009} to obtain the joint posterior distribution of the parameters of interest. Let $\pi(\cdot)$ and $p(\cdot)$ denote the prior and posterior distribution respectively.

For each individual $i$ we have a vector $\bmath{y}_i^k=(y_{i1}^k,y_{i2}^k, \dots, y_{in_{i}}^k)^T$ of $n_i$ responses of biomarker $k$ generated from equation~(\ref{eq:generic_univ}). The joint probability density function for $Y_i^{k}=(Y_{i1}^k,Y_{i2}^k,\dots,Y_{in_i}^k)^T$, conditional on parameters, is given by:
\begin{align}
f(\bmath{y}_i^k|\bmath{\beta}_i^k,\bmath{t}_{i},\tau_i,\sigma_{\epsilon^k})
& = f(y_{i1}^k,y_{i2}^k,\dots,y_{in_i}^k|\bmath{\beta}_i^k,\bmath{t}_{i},\tau_i,\sigma_{\epsilon^k})\nonumber\\
& = (2\pi)^{-\frac{n_i}{2}}\abs{\Sigma_i}^{\frac{1}{2}}e^{-\frac{1}{2}\big(\bmath{y}_i^k-\bmath{g_k}(\tau_i,\bmath{t}_i,\bmath{\beta}_i^k)\big)^T(\Sigma_i)^{-1}\big(\bmath{y}_i^k -\bmath{g_k}(\tau_i,\bmath{t}_i,\bmath{\beta}_i^k)\big)}
\end{align}
where $\Sigma_i=\sigma_{\epsilon^k}^2\bmath{I}_{n_i^k}$. \par

For univariate mixed-effects models, the generic form of the joint posterior distribution is:
\begin{equation}
\begin{split}
p(\bmath{\Theta_1} |\bmath{y}^k)\propto &\prod_{i=1}^{n} \Big\{f(\bmath{y}_{i}^k|\bmath{\beta}_i^k,\bmath{t}_{i},\tau_i,\sigma_{\epsilon^k})\pi(\bmath{\beta}_i^k|\mu_{\beta^k},\Sigma_{\beta^k}) \pi(\tau_i)\Big\} \times \pi(\mu_{\beta^k})\pi(\Sigma_{\beta^k})\pi(\sigma_{\epsilon^k})
\end{split}
\label{eq:univ_pd} 
 \end{equation} 
 where $\bmath{\Theta_1}=\big\{\bmath{\beta}_{i}^k,\mu_{\beta^k},\Sigma_{\beta^k},\sigma_{\epsilon^k},\tau_i\big\}$.

When two outcomes are measured at the same time, for each individual $i$ we have a vector $Y_i=(Y_{i1}^1,Y_{i2}^1,\dots,Y_{in_i}^1,Y_{i1}^2,Y_{i2}^2,\dots,Y_{in_i}^2)^T$. The joint probability density function for $Y_i$ conditional on parameters, can be expressed as:
\begin{equation}
\begin{split}
f(\bmath{y}_i^1,\bmath{y}_i^2|\bmath{\beta_i}^1,\bmath{\beta_i}^2,\bmath{t}_{i},\tau_i,\Sigma_{\epsilon})
=(2\pi)^{-\frac{n_i}{2}}\abs{\Sigma_{\epsilon}}^{\frac{1}{2}}e^{-\frac{1}{2}\big(Q_i^T\Sigma_{\epsilon}^{-1}Q_i\big)}
\end{split}
\end{equation}
where 
$
Q_i=\big(\bmath{y}_i^1-\bmath{g_1}(\tau_i,\bmath{t}_i,\bmath{\beta}_i^1),\bmath{y}_i^2-\bmath{g_2}(\tau_i,\bmath{t}_i,\bmath{\beta}_i^2)\big)^T
$ and
$\Sigma_{\epsilon}=\begin{pmatrix}
\sigma_{\epsilon^1}^2\bmath{I}_{n_i^1} & \bmath{0}\\
\bmath{0}&\sigma_{\epsilon^2}^2\bmath{I}_{n_i^2}
\end{pmatrix}$.

The generic form of the joint posterior distribution for a bivariate outcome is given by:
\begin{equation}
p(\bmath{\Theta_2}|\bmath{y}^1,\bmath{y}^2)\propto \prod_{i=1}^{n}
\Big\{
f(\bmath{y}_{i}^1,\bmath{y}_{i}^2|\bmath{\beta_i}^1,\bmath{\beta_i}^2,\bmath{t}_{i},\tau_i,\Sigma_{\epsilon})\pi(\tau_i)
\pi(\bmath{\beta}_i^1,\bmath{\beta}_i^2|\mu_{\bmath{\beta}},\Sigma_{\bmath{\beta}})\Big\}\pi(\mu_{\bmath{\beta}})\pi(\Sigma_{\bmath{\beta}})\pi(\Sigma_{\epsilon})
\label{eq:joint_pd}
\end{equation}
where $\bmath{\Theta_2}=\{\bmath{\beta_{i}}^1,\bmath{\beta_{i}}^2,\mu_{\bmath{\beta}},\Sigma_{\bmath{\beta}},\Sigma_{\epsilon},\tau_i \}$.

\begin{figure}
\centerline{
\includegraphics[width=8cm,height=7.5cm]{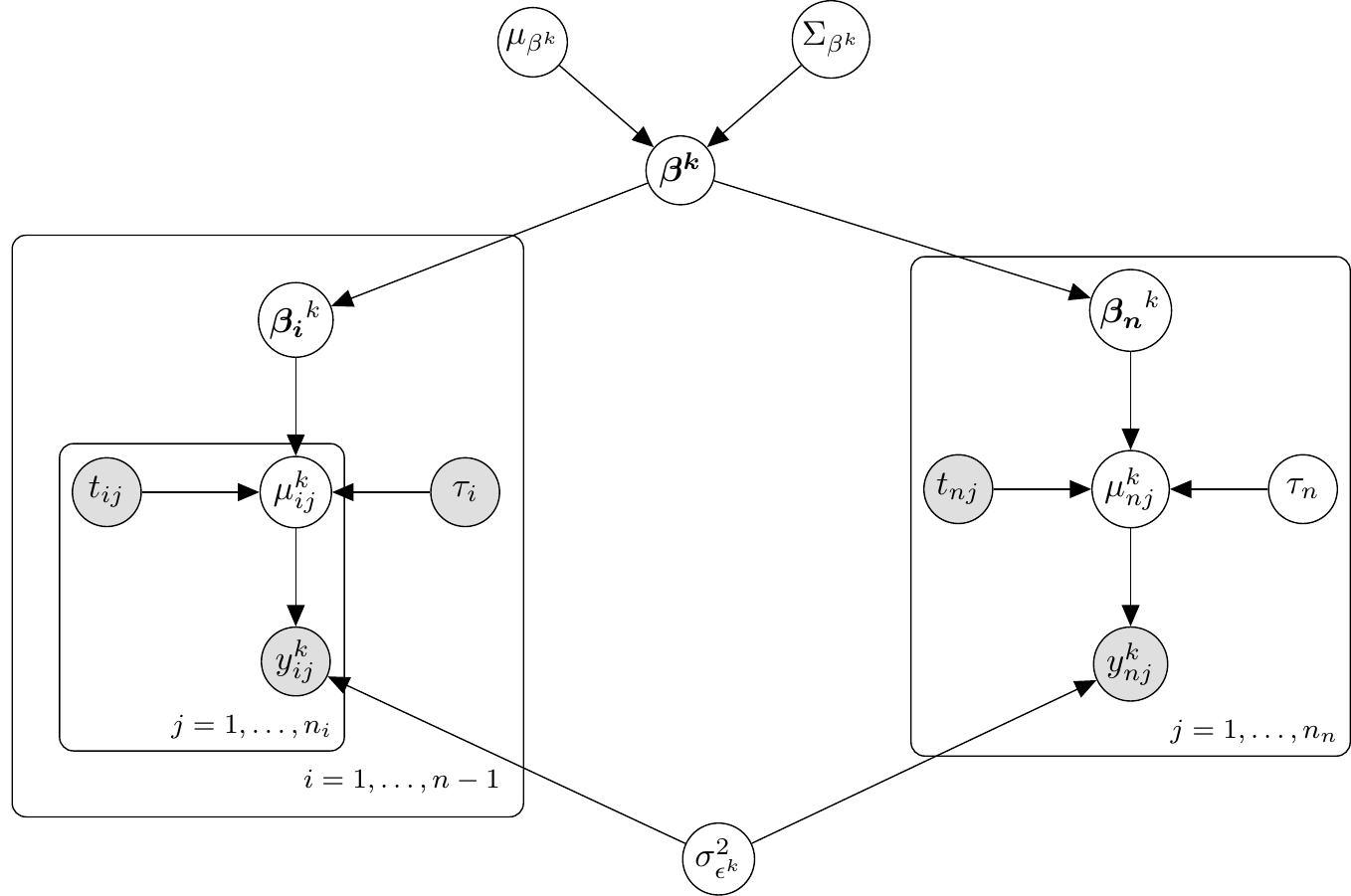}}
\caption[Example DAG of mixed models]{Directed acyclic graph depicting the univariate model for the $k^{th}$ biomarker as shown in equation (\ref{eq:generic_univ}) by assuming that the first $n-1$ are the in-sample individuals and the $n^{th}$ is the new individual. Data are represented by shaded circles and the rest of the nodes are unknown random variables.}
\label{fig:DAG_univ}
\end{figure}

For each simulated dataset, the first 100 individuals are considered
as the \enquote{in-sample} individuals for whom the seroconversion
time ($\tau_{1:(n-1)}$) is known. The analysis is conducted by
assuming that we have only one new individual ($n=101$) in each
dataset, with unknown seroconversion time, as shown in
Figure~\ref{fig:DAG_univ}. 

The joint posterior
distribution of the univariate model displayed in
Figure~\ref{fig:DAG_univ} is therefore: 
\begin{align}
p(\tau_{n},\boldsymbol{\Theta}|\boldsymbol{y}^k,\boldsymbol{t},\tau_{1:(n-1)})
& \propto \prod_{i=1}^{n-1}\Big\{ f(\boldsymbol{y}_{i}^k|\boldsymbol{\beta_i}^k,\boldsymbol{t}_{i},\tau_i,\sigma_{\epsilon^k}^2)
\pi(\boldsymbol{\beta}_i^k|\mu_{\boldsymbol{\beta^k}},\Sigma_{\boldsymbol{\beta^k}})\Big\}\nonumber\\
& \times
f(\boldsymbol{y}_{n}^k|\boldsymbol{\beta_n}^k,\boldsymbol{t}_{n},\tau_n,\sigma_{\epsilon^k}^2)\pi(\boldsymbol{\beta}_n^k|\mu_{\boldsymbol{\beta^k}},\Sigma_{\boldsymbol{\beta^k}})
\pi(\tau_{n}) \pi(\mu_{\beta^k})\pi(\Sigma_{\beta^k})\pi(\sigma_{\epsilon^k})
\end{align}
where
$\boldsymbol{\Theta}=\{\boldsymbol{\beta}_{i}^k,\mu_{\beta^k},\Sigma_{\beta^k},\sigma_{\epsilon^k}^2\}$. Similarly,
the joint posterior distribution for a bivariate outcome is obtained
by replacing $\boldsymbol{y}^k$ with
$\boldsymbol{y}^1,\boldsymbol{y}^2$ and
$\boldsymbol{\Theta}=\{\boldsymbol{\beta}_{i}^1,\boldsymbol{\beta}_{i}^2,\mu_{\boldsymbol{\beta}},\Sigma_{\boldsymbol{\beta}},\Sigma_{\epsilon}\}$.

\subsection*{Priors}
Each new individual is assumed
    to have unknown seroconversion time occurring in their
    seroconversion interval, $\tau_{n} \in [t_n^{-ve},t_n^{+ve}]$. Our
    {\it a priori} belief is therefore that $\tau_{n}\sim
    Uniform(0,1)$, since each seroconversion interval is of length 1 year. Vague Gaussian priors, $N(0,10^6)$, are placed on the means of the random effects $(\mu_{\beta^k},\mu_{\boldsymbol{\beta}})$, while $\sigma_{\epsilon^{k}}^2$ is given an inverse-Gamma prior, $\pi(\sigma_{\epsilon^{k}}^2)\sim IG(2,0.01)$.  Each variance-covariance matrix of the random effects ($\Sigma_{\beta^{k}}, \Sigma_{\boldsymbol{\beta}}$) and of the measurement error ($\Sigma_{\epsilon}$), is given an inverse-Wishart prior with degrees of freedom equal to one plus the matrix dimension. These priors effectively place a uniform distribution on each of the correlation parameters \citep{gelman2007}.

For each simulated dataset, MCMC samples
from the marginal posterior distribution of the seroconversion time
$p(\tau_{n}|\boldsymbol{y},\boldsymbol{t},\tau_{1:(n-1)})$ are obtained. These
samples are used to derive posterior probabilities, $P_X = Pr(\tau_{n} \leq
X|\boldsymbol{y},\boldsymbol{t},\tau_{1:(n-1)})$, of a new individual
having seroconverted in the last $X$ years before HIV diagnosis. We
evaluate the predictive ability of the proposed models in quantifying
the recency of HIV infection by calculating these probabilities for $X = 0.167, 0.333$ and $0.5$ years before HIV diagnosis ($P_2, P_4$ and $P_6$
respectively, corresponding to 2, 4 and 6 months).

\section{Results}{\label{sec:results}}
\subsection{Probability of recent seroconversion}
The simulated data are initially analysed assuming
that the new individual contributes only a single measurement taken at
HIV diagnosis. The analysis is repeated including
consecutive measurements of the new individual taken either two weeks
or one month after HIV diagnosis. No significant differences are
observed when we add consecutive measurements to the estimation process. We therefore present results based only on the measurements taken at HIV diagnosis.

The distributions of the probabilities $P_2, P_4, P_6$ are summarized over
the 100 simulated datasets in
Figure~\ref{fig:probsc_Xmonths_prior_real246} for each of the 5 \enquote{out-of-sample} individuals. For a perfectly discriminatory biomarker, we would expect these probabilities to lie near 0 or 1 depending on the truth. For instance, $P_2$ should lie around 0 for all patients with true seroconversion occurring more than two months before HIV diagnosis.

\subsubsection{Single outcome}
The linearly evolving biomarker AR1 leads to very similar estimates of $P_2,P_4$ and $P_6$ for each new patient. In particular, $P_2$ is estimated to lie below 0.05 even for a new patient with true seroconversion occurring five days before HIV diagnosis (see Figure~\ref{fig:probsc_Xmonths_prior_real246}). A possible explanation might be that biomarker AR1 is only gradually evolving, so that its observed values are too similar across time.

The non-linear biomarkers of antibody
response with a low asymptote, such as AR2
and AR3, perform worse compared to AR4 (see
Web Appendix C, Figure 8). They lead to flat posterior distributions
of seroconversion time (see Web Appendix B Figures 3-7). On the other
hand, AR4 and viral load seem to be quite
discriminative, providing strong information on seroconversion
  time, especially for seroconversions occurring a few days before HIV diagnosis. However, as both biomarkers approach their asymptotes, their ability to discriminate the seroconversion time vanishes. For instance, $P_6$ is greater than 0.6 for patients with long-standing infections, using the univariate model of viral load.
  
\subsubsection{Bivariate outcomes}
\begin{figure}
     \begin{subfigure}[b]{0.99\textwidth}
    \includegraphics[width=17cm,height=6.8cm]{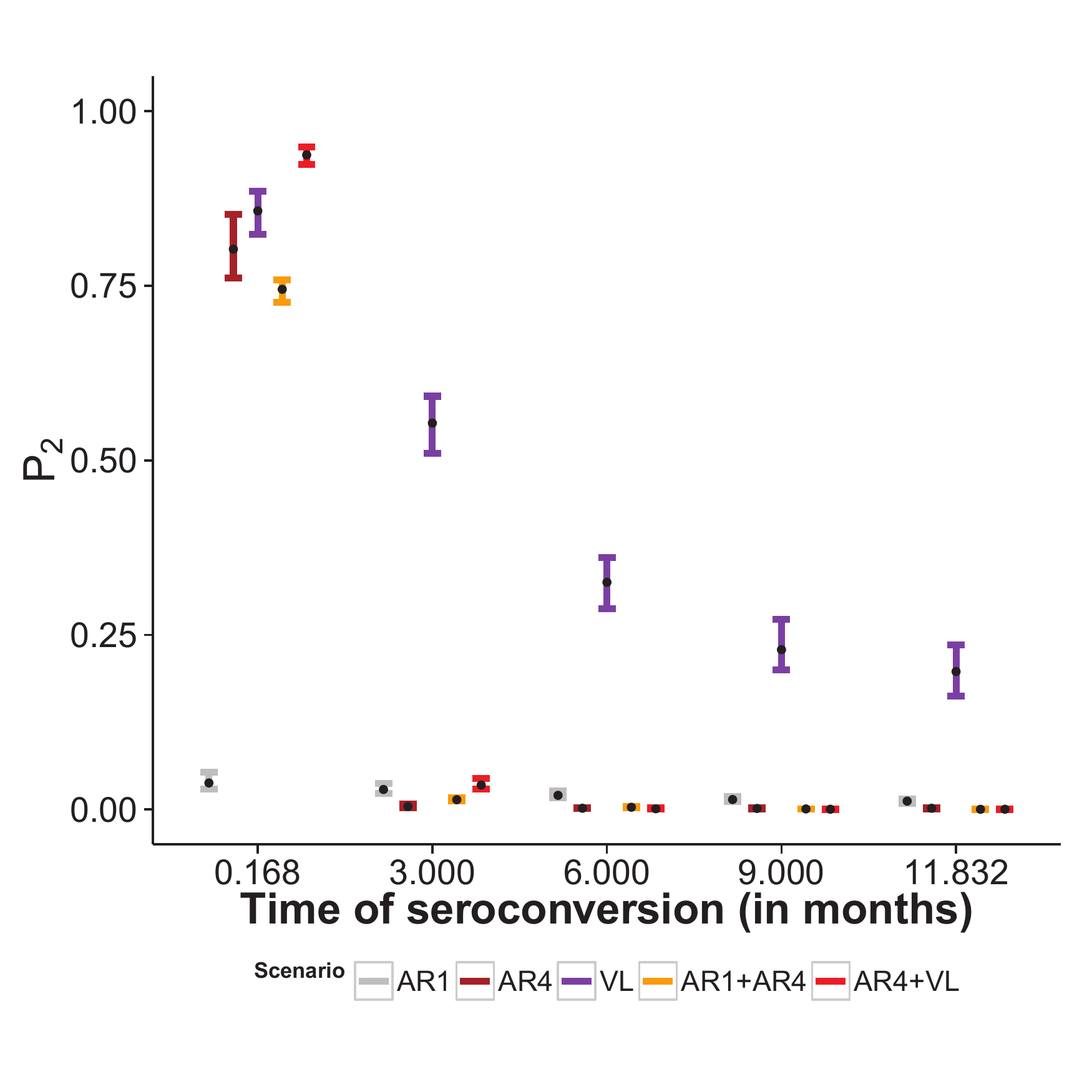} 
  \end{subfigure} \\ \vspace{0.2em}
  \begin{subfigure}[b]{0.99\textwidth}
    \includegraphics[width=17cm,height=6.8cm]{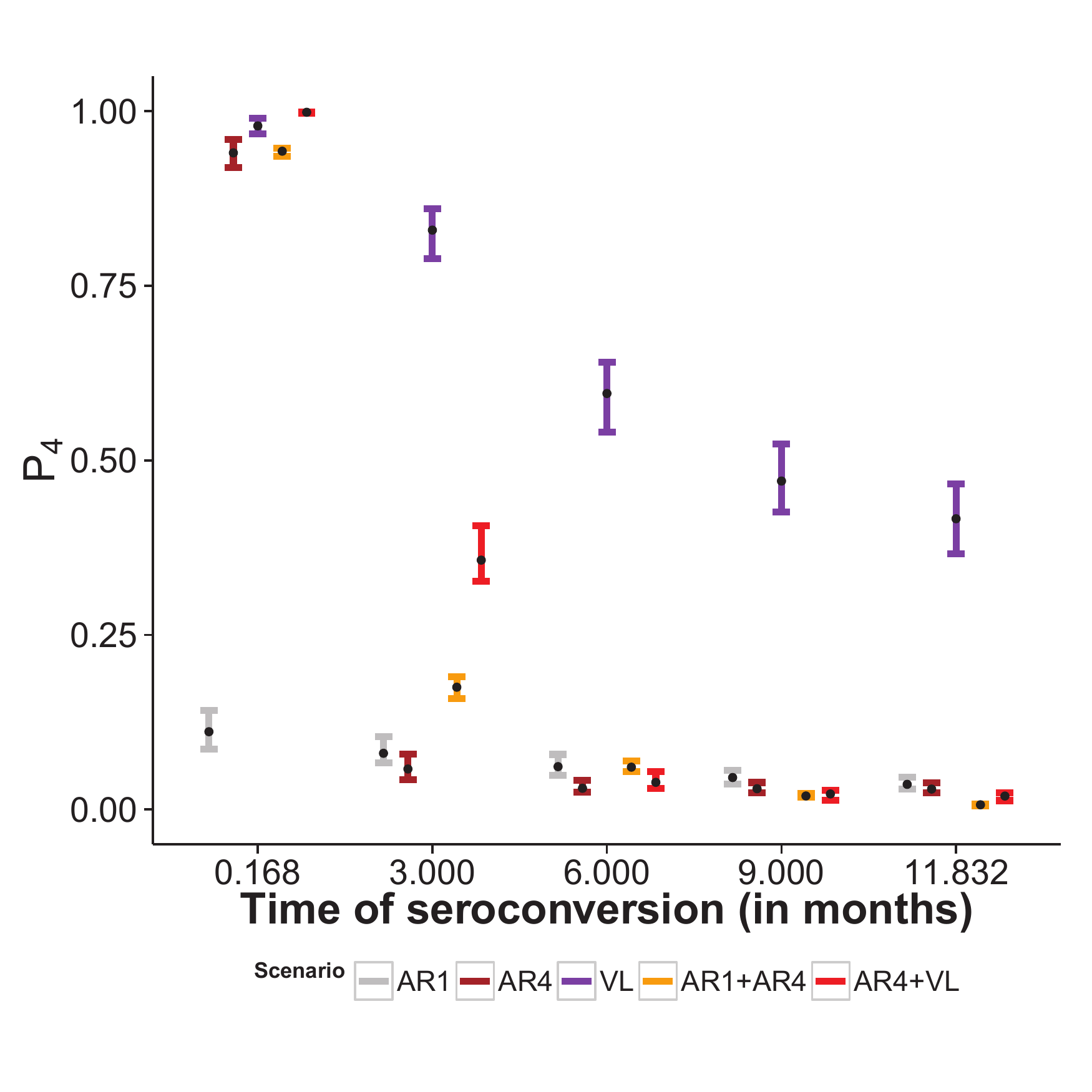} 
  \end{subfigure} \\ \vspace{0.2em}
   \begin{subfigure}[b]{0.99\textwidth}
    \includegraphics[width=17cm,height=6.8cm]{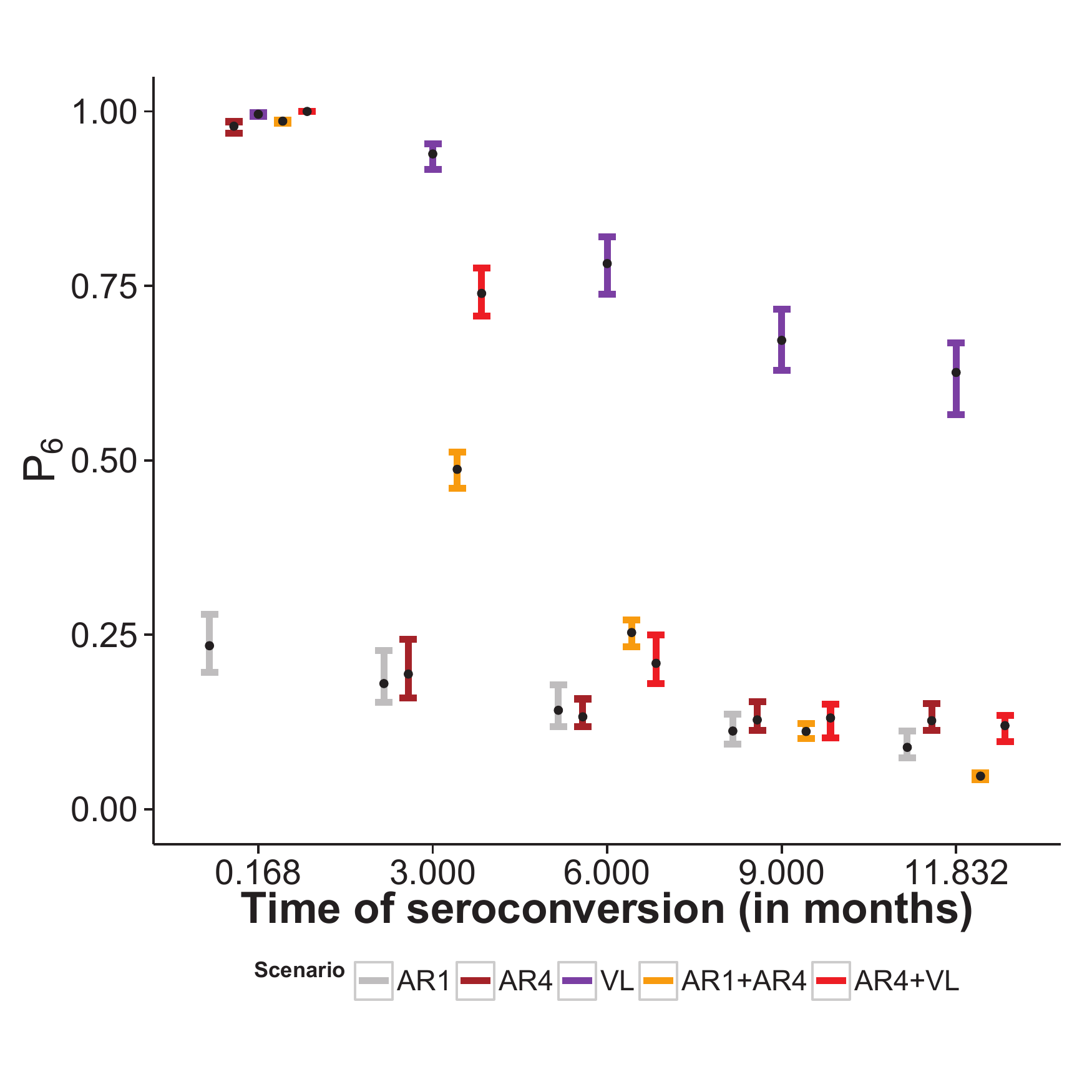} 
  \end{subfigure} 
\caption{Probability of having seroconverted in the last two, four, and six months before HIV diagnosis under the realistic scenario. The black dot represents the median and the upper and lower limits of the error bars show the $25^{th}$ and $75^{th}$ percentile of the distribution over the simulated datasets respectively.}         
  \label{fig:probsc_Xmonths_prior_real246}  
\end{figure}

The quantification of recency is improved by using joint models of two
biomarkers of interest. In particular, the joint model of
AR4 and VL is able to
accurately estimate $P_2$ for all new patients. It leads to estimates
above 0.9 and below 0.1 for very recent and
long-standing infections respectively. Similar results are obtained
when we use a bivariate outcome of two antibody-response
biomarkers. Notably, the combination of AR4
and AR1 leads to the most accurate estimates of 
  $P_2,P_4$ and $P_6$ for those individuals with long-standing
infections (see Figure~\ref{fig:probsc_Xmonths_prior_real246}). More
specifically, it gives estimates of $P_6$ below
0.05 for a new patient with $\tau_{n}=0.986$ years before HIV
diagnosis, compared to 0.2 when a univariate model of
AR4 is used. This improvement might be due to the fact that AR1 is linear and so does not plateau, providing some information on recency even if HIV diagnosis takes place a long time after infection.

The accuracy of the estimation clearly depends on particular
characteristics of biomarkers of interest, as well as on the timing of
the first measurement. If we had to choose only a single biomarker to
quantify recency, we would prefer a rapidly evolving biomarker such as
AR4. If viral load is available, a joint model of antibody response
and viral load will lead to more accurate estimates of the probability
of having seroconverted recently. However, even a joint model of two biomarkers lacks the ability to provide reliable estimates of the recency for all new patients.

\subsection{Ideal Scenario}
A further simulation exercise (see Web Appendix A) reveals that the
  magnitude of between-individual heterogeneity has a significant
  effect on the discriminatory ability of each biomarker or combination
  of biomarkers. To investigate this effect, we generated and analysed data as shown previously, but with the variance of all the random effects being set to 0.01, in an ``\textbf{ideal scenario}''.

\begin{figure}
     \begin{subfigure}[b]{0.99\textwidth}
    \includegraphics[width=17cm,height=6.8cm]{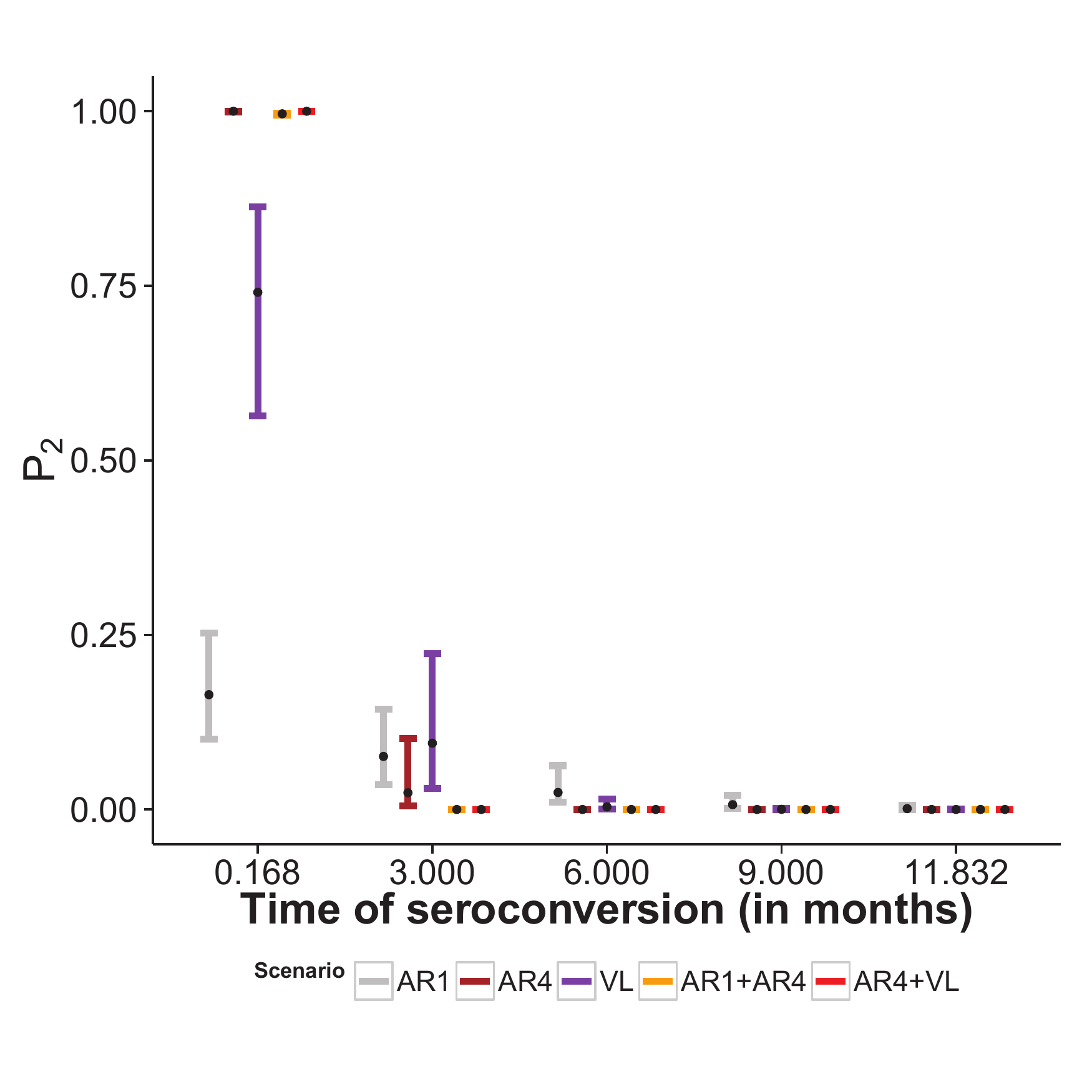} 
  \end{subfigure} \\ \vspace{0.2em}
  \begin{subfigure}[b]{0.99\textwidth}
    \includegraphics[width=17cm,height=6.8cm]{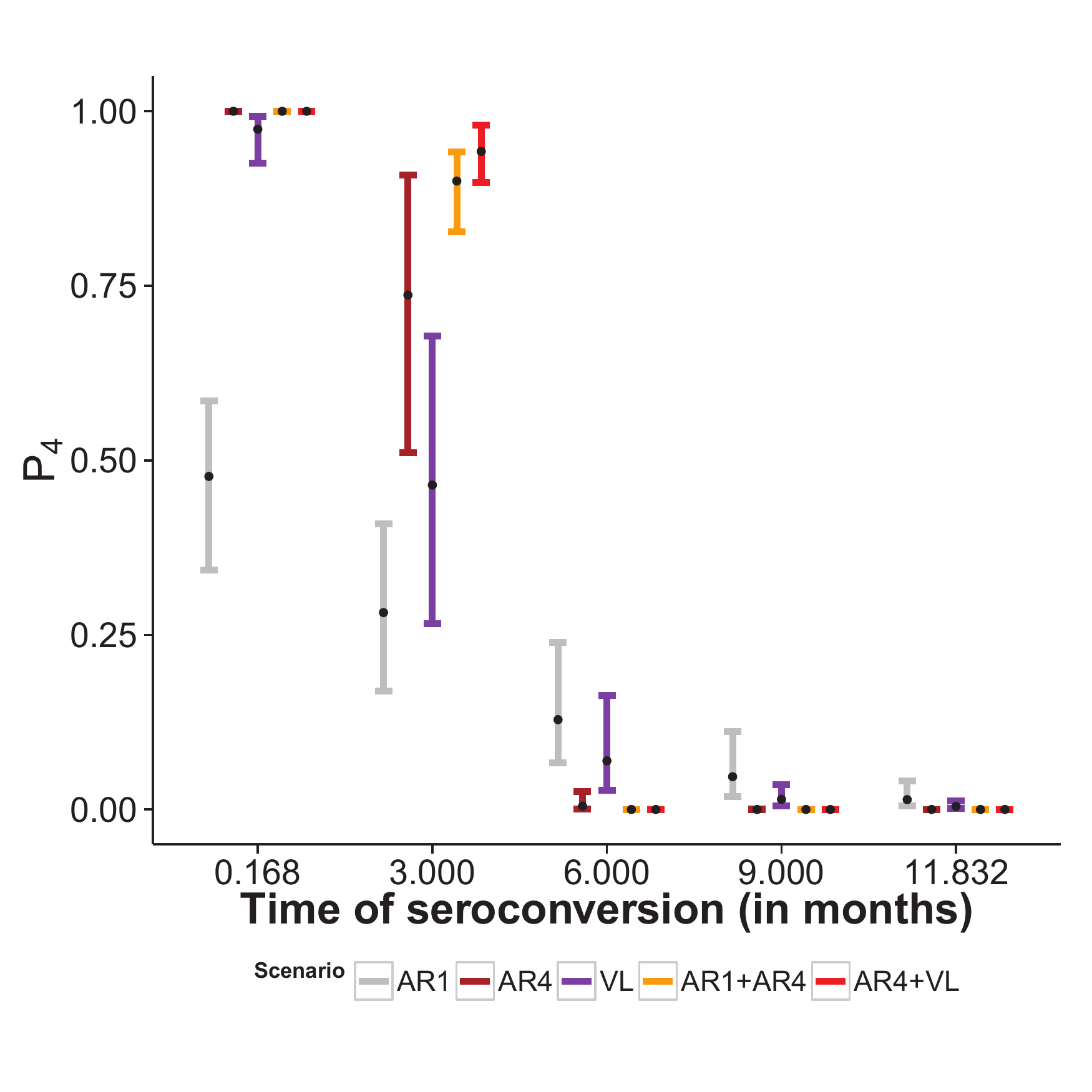} 
  \end{subfigure} \\ \vspace{0.2em}
   \begin{subfigure}[b]{0.99\textwidth}
    \includegraphics[width=17cm,height=6.8cm]{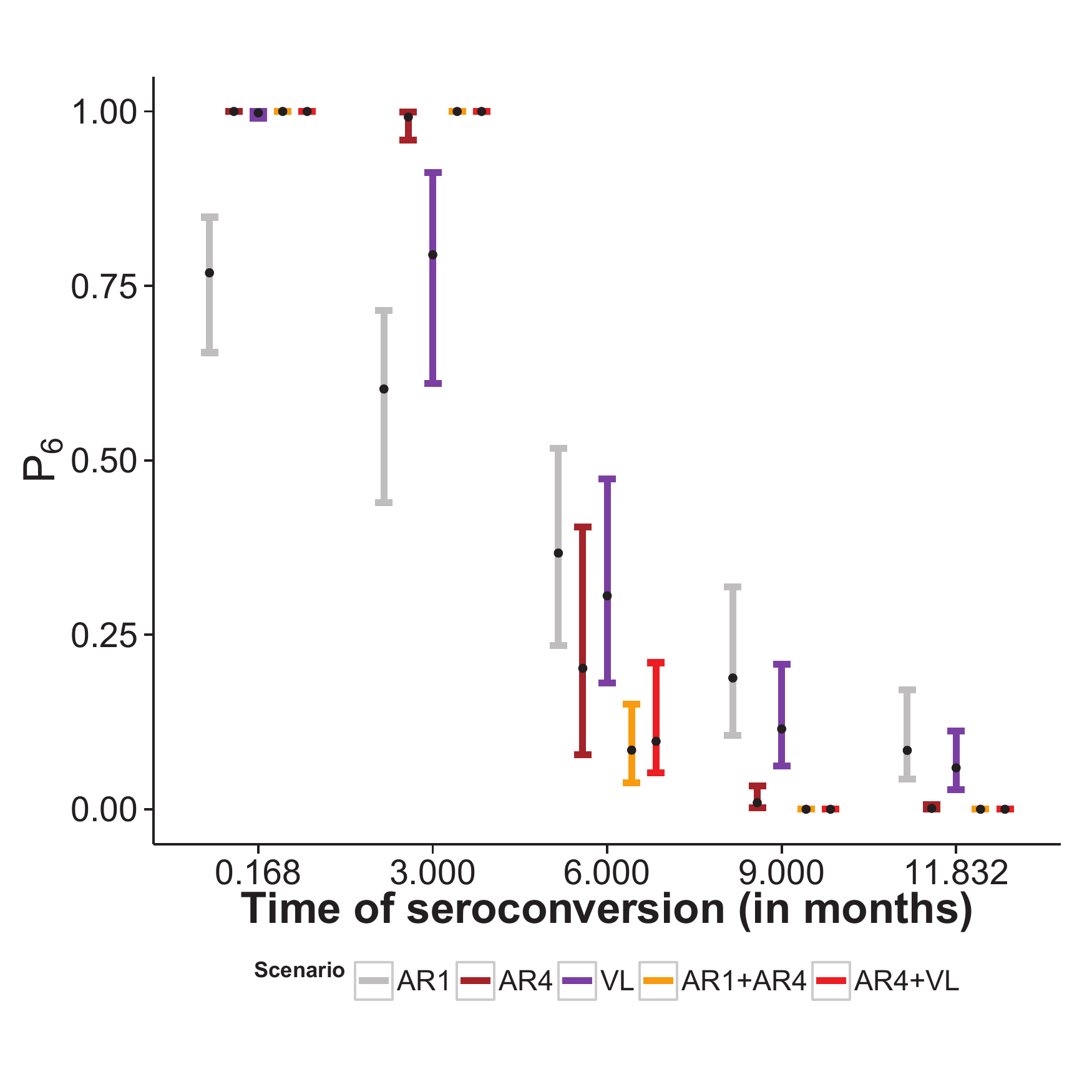} 
  \end{subfigure} 
 \caption{Probability of having seroconverted in the last two, four, and six months before HIV diagnosis under the ideal scenario. The black dot represents the median and the upper and lower limits of the error bars show the $25^{th}$ and $75^{th}$ percentile of the distribution over the simulated datasets respectively.}        
  \label{fig:probsc_Xmonths_prior_ideal246}  
\end{figure}

We obtain more accurate estimates of the probabilities $P_2,P_4$ and $P_6$ for any new individual (see Figure~\ref{fig:probsc_Xmonths_prior_ideal246}), when univariate and bivariate outcomes are generated under the ideal compared to the realistic scenario.

\subsubsection{Single outcome}
Univariate models of AR4 and
viral load lead to high values of $P_2$ for recently infected
individuals, and very low values for long-standing infections. The same pattern is observed for $P_4$ when univariate outcomes are used. However, for a patient with true seroconversion occurring exactly six months before HIV diagnosis, all univariate outcomes lead to $P_6$ below 0.6 (See Web Appendix Figure 9).

\subsubsection{Bivariate outcomes}
Bivariate joint models improve the estimates of $P_2,P_4$ and $P_6$
compared to their univariate counterparts. In particular, for a new
patient who has seroconverted a few days ($\tau_{n}=0.014$ years)
before HIV diagnosis, the bivariate joint models lead to $P_2\geq 0.95$. Furthermore,
for patients who have seroconverted more than two months before HIV
diagnosis, both bivariate models lead to estimates of $P_2\leq 0.05$.

For $\tau_{n}=0.5$, all models give a small probability of having seroconverted in the last six months, with the joint models leading to the smallest estimates. A possible explanation might be that the non-linear biomarkers of antibody response usually approach their asymptote around the first six months from seroconversion \citep{chawla2007,hargrove2008,suligoi2011,sweeting2010}. Therefore, all measurements taken six months after seroconversion are very similar and are indicative of long-standing infections, but cannot discriminate the actual time of seroconversion.

 \begin{figure}
 \captionsetup[subfigure]{justification=centering}
  \begin{subfigure}[b]{0.5\linewidth}
    \centering
    \includegraphics[width=4.5cm,height=6.3cm]{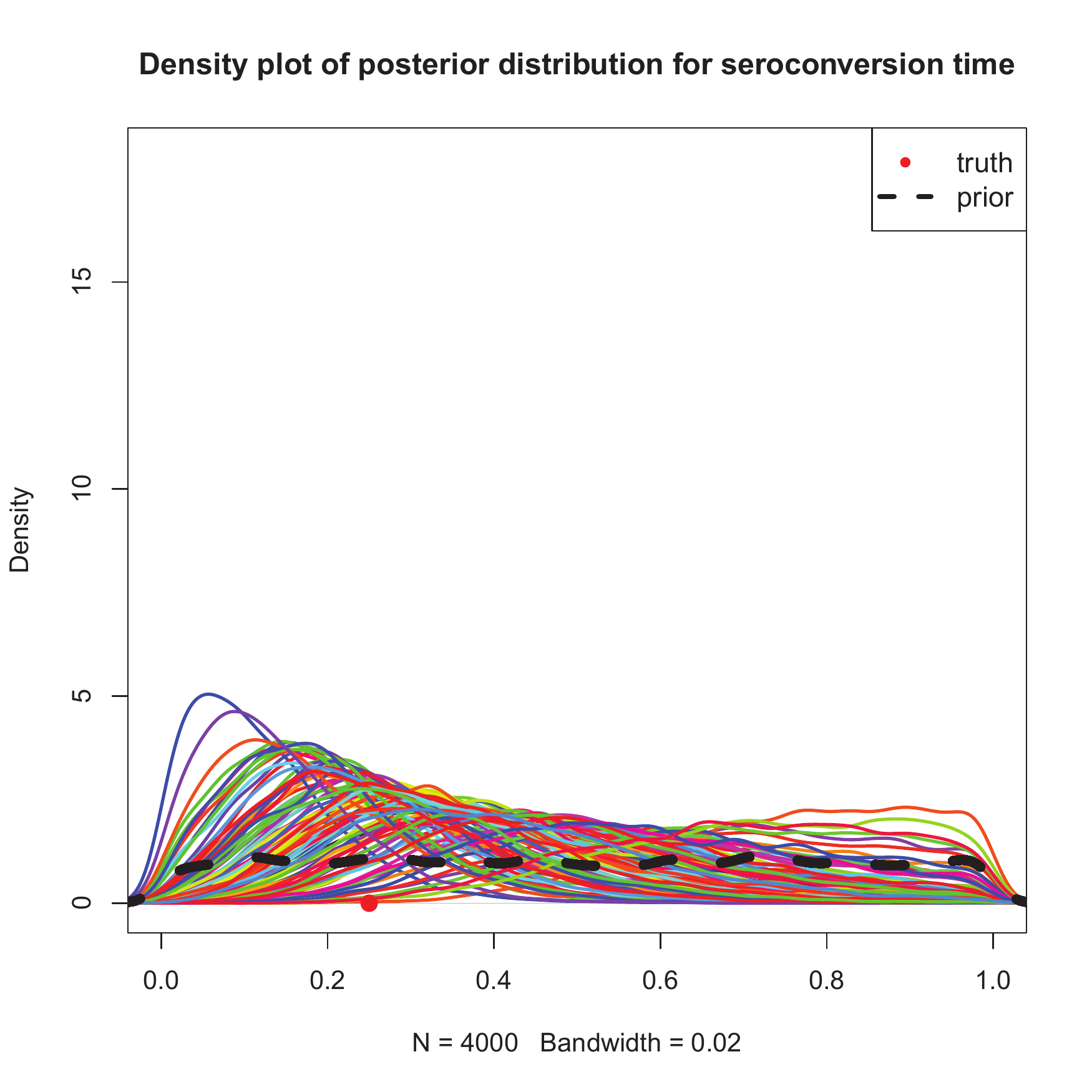} 
       \caption{VL ideal scenario}
       \label{fig:vlideal_pat2}
  \end{subfigure} 
  \begin{subfigure}[b]{0.5\linewidth}
    \centering
    \includegraphics[width=4.5cm,height=6.3cm]{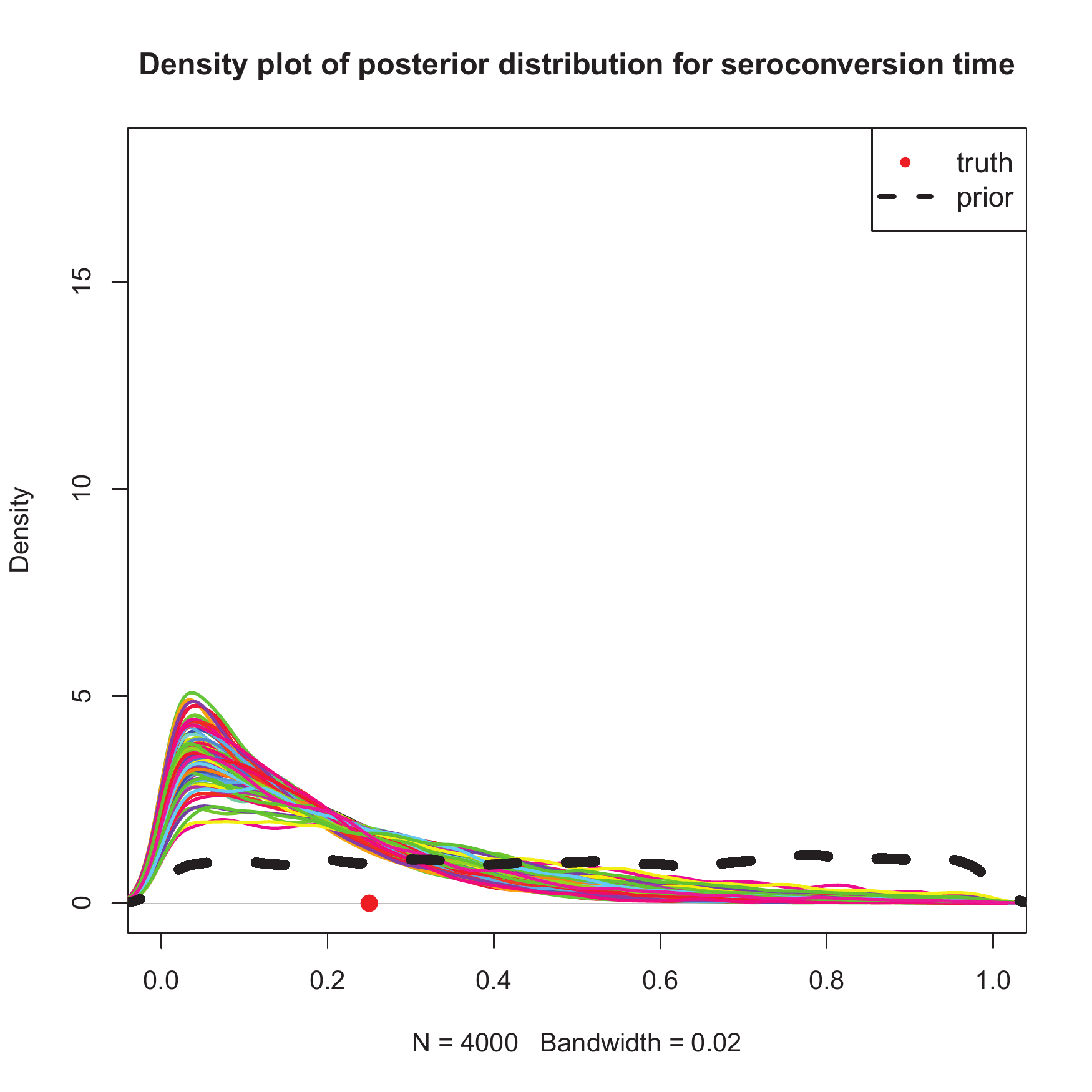} 
       \caption{VL realistic scenario}
       \label{fig:vlrealistic_pat2}
  \end{subfigure} \\
  \begin{subfigure}[b]{0.5\linewidth}
    \centering
    \includegraphics[width=4.5cm,height=6.3cm]{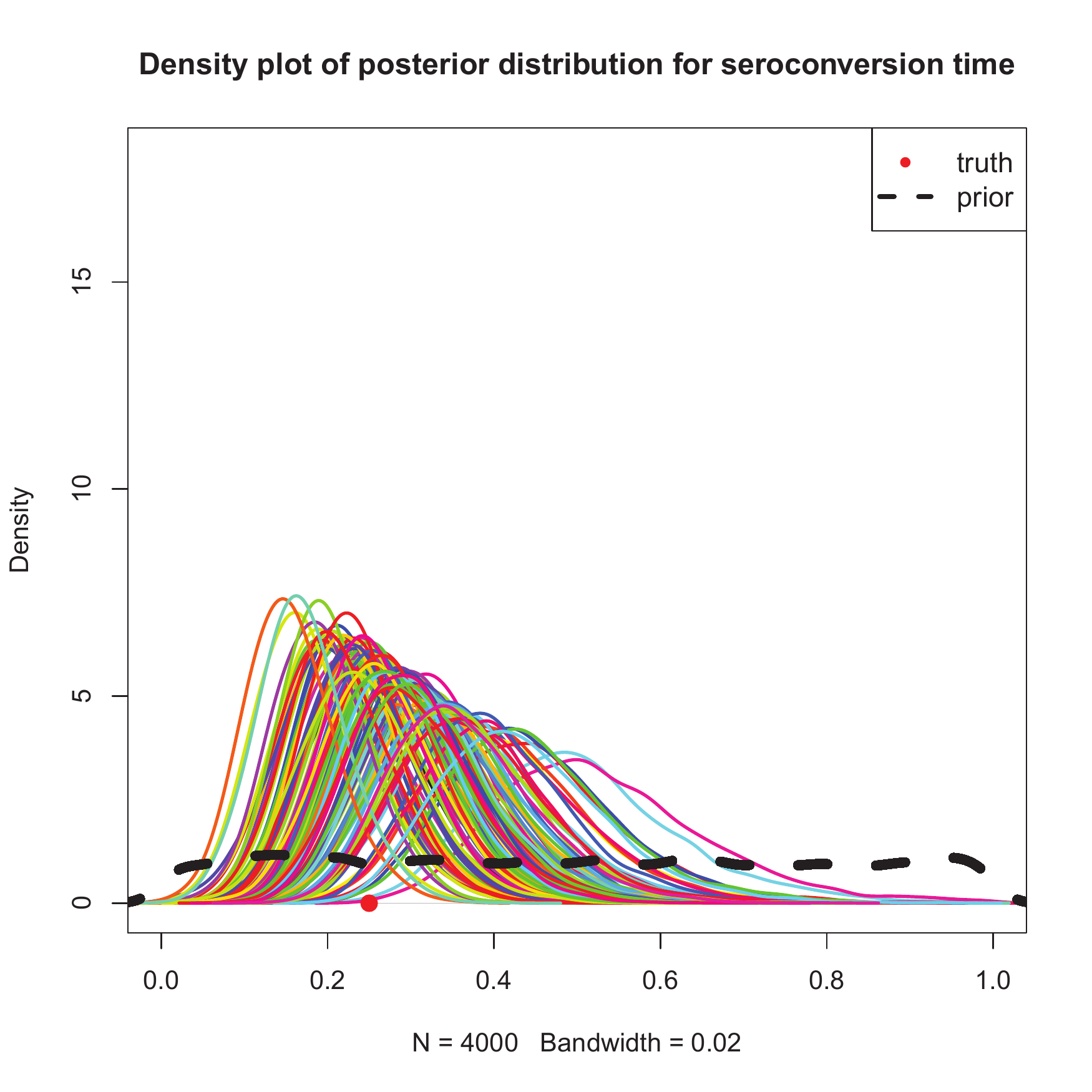} 
       \caption{AR4 ideal scenario}
       \label{fig:ar4idealvl_pat2}
  \end{subfigure} 
  \begin{subfigure}[b]{0.5\linewidth}
    \centering
    \includegraphics[width=4.5cm,height=6.3cm]{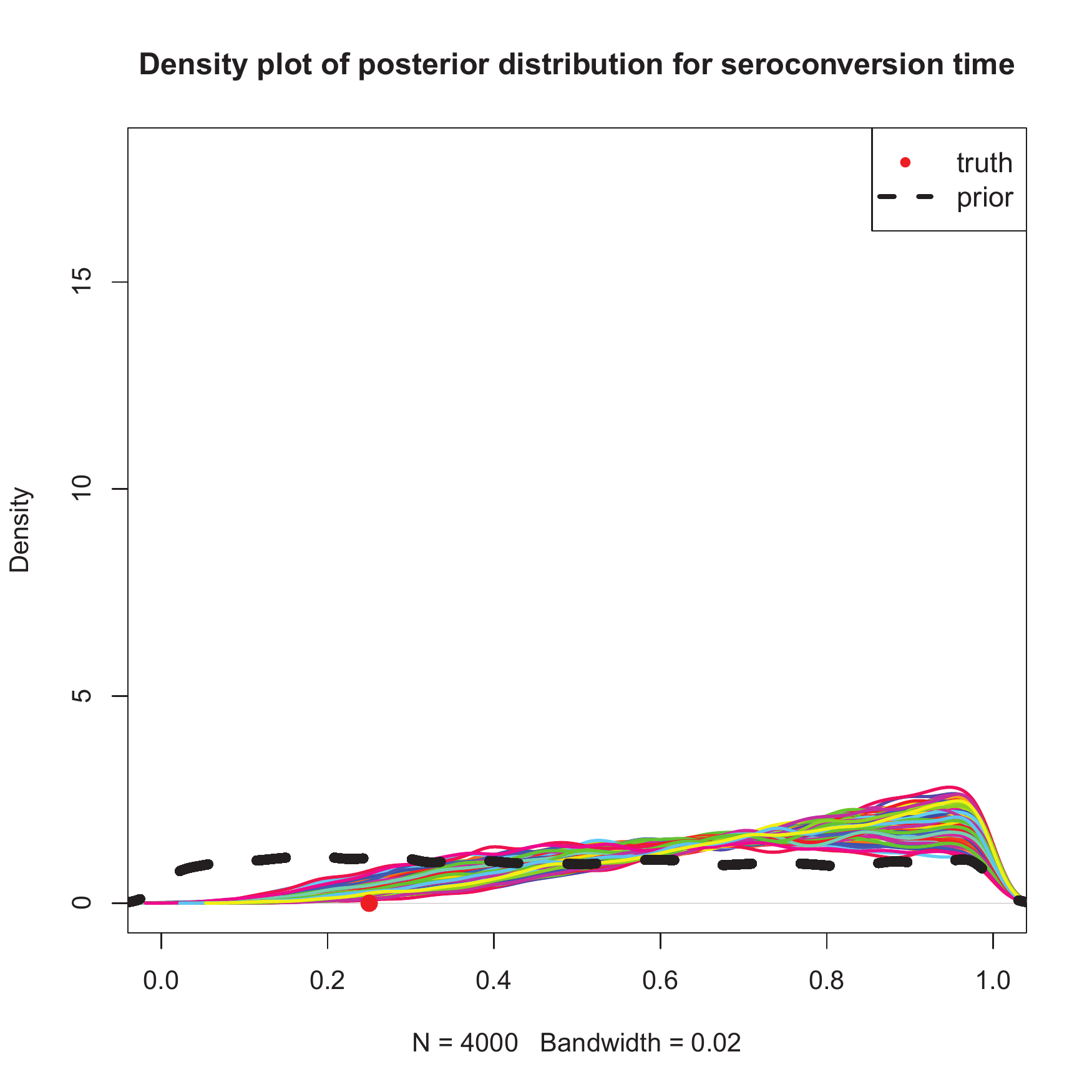} 
       \caption{AR4 realistic scenario}
       \label{fig:ar4realisticvl_pat2}
  \end{subfigure} 
  \begin{subfigure}[b]{0.5\linewidth}
    \centering
    \includegraphics[width=4.5cm,height=6.3cm]{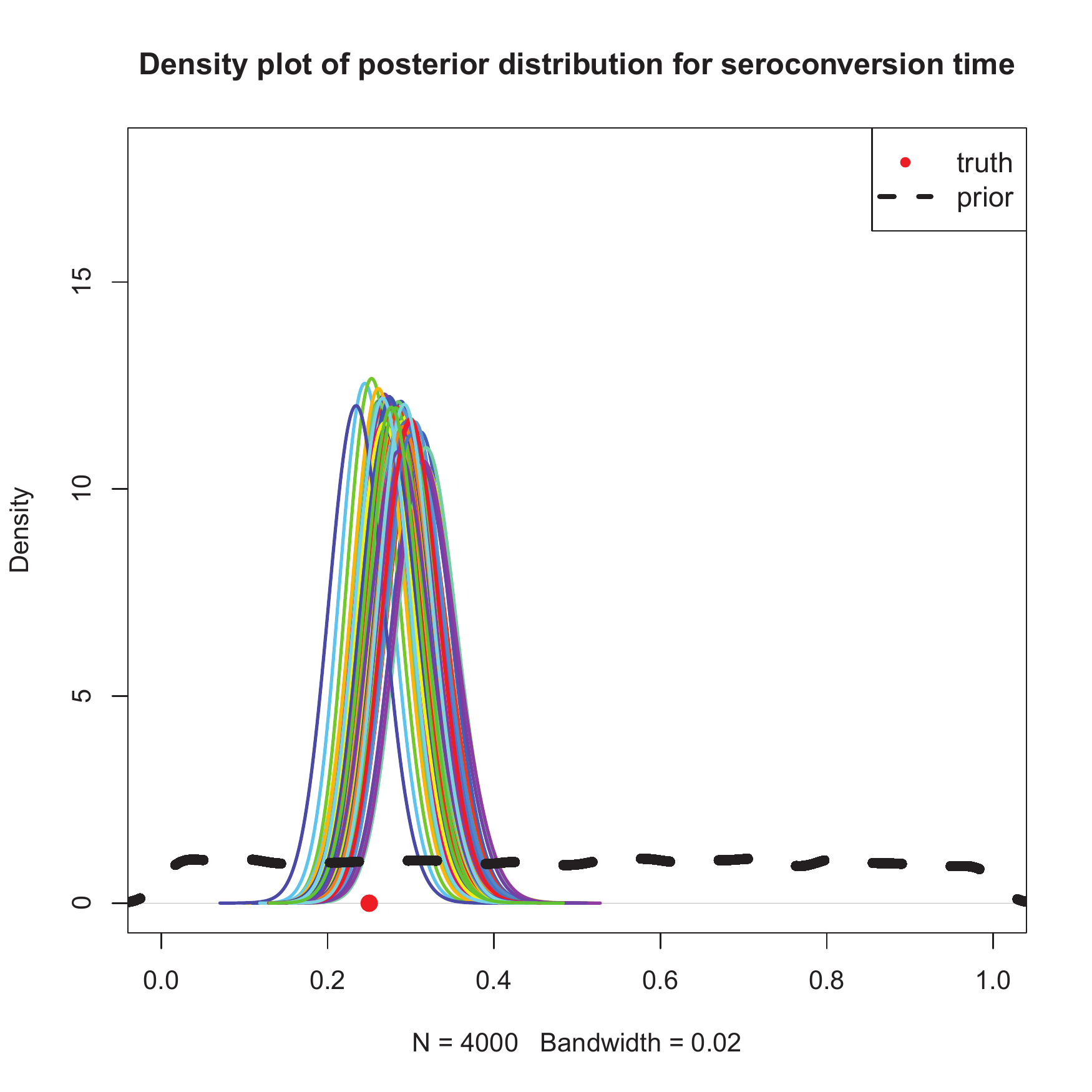} 
       \caption{VL \& AR4 ideal scenario}
       \label{fig:vlar4ideal_pat2}
  \end{subfigure} 
  \begin{subfigure}[b]{0.5\linewidth}
    \centering
    \includegraphics[width=4.5cm,height=6.3cm]{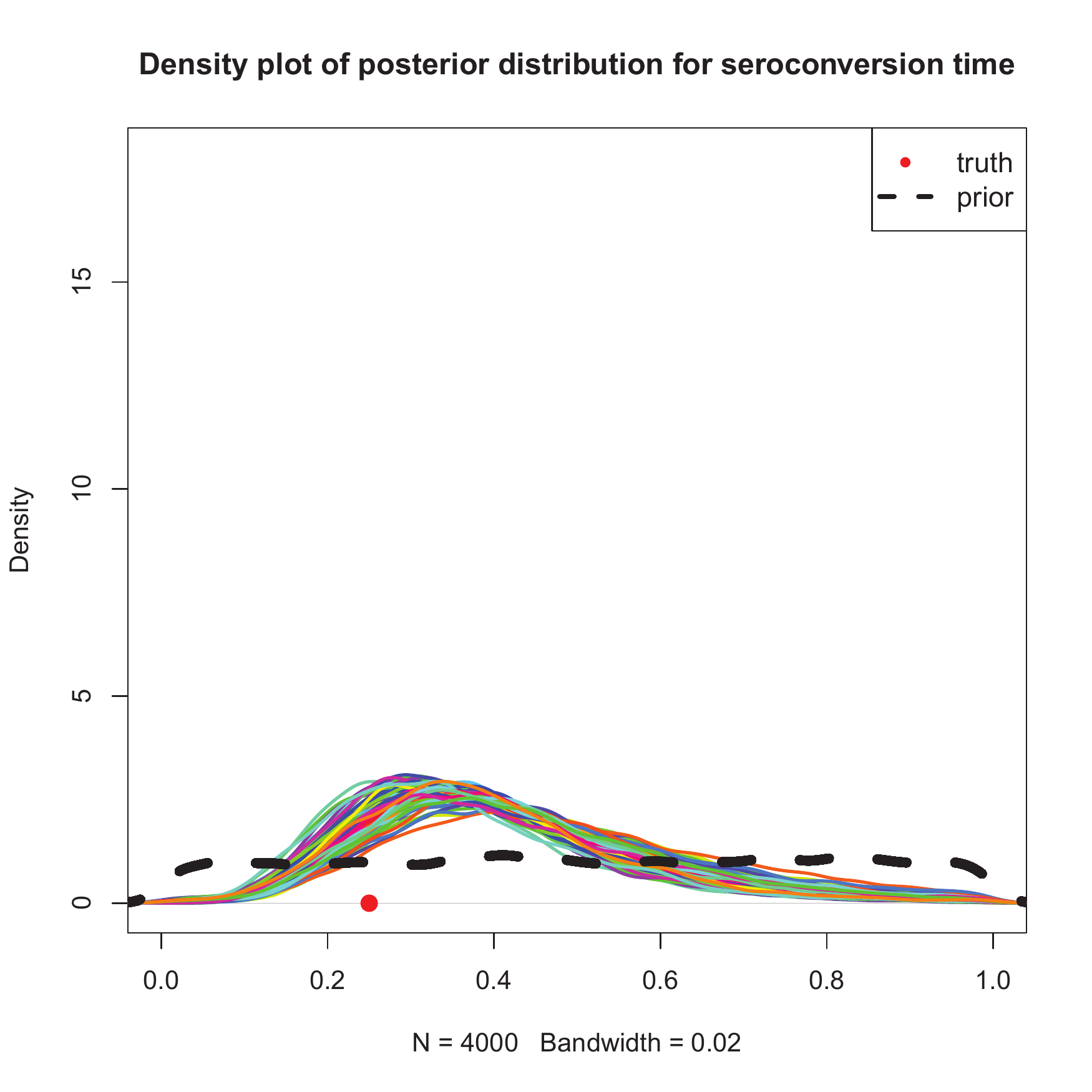} 
       \caption{VL \& AR4 realistic \\scenario}
       \label{fig:vlar4realistic_pat2}
  \end{subfigure} 
  \caption{Density plots of the posterior distribution of seroconversion time over the simulated datasets, when we use the antibody response biomarker AR4, the viral load, and their joint distribution.}   
  \label{fig:pos_vlarbio_pat2} 
\end{figure}

\section{Discussion}{\label{sec:discussion}}
We have investigated a fully Bayesian
approach to quantify the recency of HIV infection for a newly diagnosed individual, using values
of one or more biomarkers and information on biomarker evolution
obtained from a panel of HIV-infected individuals. This is the first
study to investigate the ability of biomarkers of both antibody response
and viral presence in quantifying recency at an individual level. We have also explored the characteristics that affect the discriminatory ability of such biomarkers to provide reliable estimates of the probability of having recently seroconverted. Linear and non-linear mixed-effects models are used to describe the growth/decline trajectories of biomarkers. We have introduced a bivariate non-linear mixed-effects model which allows for different non-linear trends to be modelled simultaneously.

To our knowledge, few studies with their main interest being the
estimation of the seroconversion time have been published, and usually
the number of CD4 T-cell counts is used as a biomarker of interest
\citep{munoz1989,munoz1992,dubin1994}. Bivariate linear mixed-effects
models have been proposed for markers of immunological and virological
status \citep{chakraborty2003,pantazis2005,thiebaut2005} but examples
of multivariate non-linear mixed-models are less common in the context
of HIV \citep{sommen2011,konikoff2015}. The bivariate non-linear
mixed-effects models proposed in Section~\ref{sec:methods} can be used
when the aim of the study is to explore the association between the
evolutions of two non-linear outcomes. The proposed method can be
easily extended to a multivariate non-linear model if more than two
outcomes are available.

The results of the simulation study suggest that we are able to learn
about the probability of having recently seroconverted from
longitudinal data on biomarkers of recent infection. The accuracy of
the estimation is highly influenced by particular characteristics of
markers, as well as the time of HIV diagnosis. The magnitude of the growth or decline rate plays
a vital role in the estimation, with rapidly-evolving biomarkers
(e.g. 3-6 months) providing more precise estimates of
recency. The results indicate also that
the level of the asymptote of the non-linear biomarkers affects their ability to discriminate the
  recency of infection. 

In practice, physicians are interested in using a single biomarker to
quantify the recency of HIV infection, especially if multiple biomarker measures are challenging to obtain due to time and cost
restrictions. Biomarker AR4 seems to provide reliable estimates of the
probability of having recently seroconverted when used in a univariate
model. As shown in Web Appendix A: Figure 1, the distribution of AR4 at different time points overlaps less
compared to all the other single biomarkers. Therefore, we suggest
using an antibody-response biomarker similar to
AR4, such as LAg Avidity, if restricted to a single biomarker.

However, we have demonstrated that the use of bivariate joint models improves the
quantification of recency. The resulting posterior distributions of
seroconversion time for each new individual that have narrower 95\% highest
  posterior density (HPD) intervals compared to the univariate
models, having accounted for the correlation between biomarkers. A combination of antibody
response and viral load seems to perform slightly better for those
seroconverting up to six months before HIV diagnosis. By contrast, for
seroconversions occurring nine months or almost one year before HIV
diagnosis, we obtain marginally better estimates when two
antibody-response biomarkers are used in the estimation process. This
result may be due to both non-linear
antibody-response biomarkers and viral load approaching their
asymptotes six months after seroconversion, when they are no longer discriminative. On the other hand, the antibody-response biomarker AR1, which is linearly evolving, allows the bivariate joint model to discriminate the seroconversion time for long-standing infections. Overall, we recommend a combination of two antibody-response biomarkers with different growth patterns, such as AR1 and AR4, to quantify recency of HIV infection.  

Surprisingly, no significant differences were found in the
quantification of recency when we use additional bivariate outcomes
taken two weeks or one month after HIV diagnosis for each new
individual. It seems that only the first measurement of the bivariate
outcome is adequate to distinguish recency,
especially if it is taken soon after seroconversion. On the other
hand, the univariate models become slightly more discriminative with additional information taken every two weeks after diagnosis (results not shown). 

The crucial finding to emerge from the analysis is that the
heterogeneity between individuals plays a vital role in the
discriminatory ability of biomarkers of interest. When the between-subjects variability is reduced
to a minimum, the values of the bivariate outcome are indicative of
the seroconversion time. As we increase the between-subjects
variability, the biomarkers of recency become less discriminative,
leading to very flat posterior distributions of seroconversion
time. However, it is challenging to find currently existing
  biomarkers that are as homogeneous as those generated under the
  ideal scenario. Researchers developing new and/or alternative
  biomarkers for recent infection should aim to find markers that
  have minimum heterogeneity, if they are to be valuable for estimation at individual
  level. 

The proposed method is based on specific assumptions about the number
and time of measurements, the length of seroconversion intervals,
the distribution of random effects and our prior beliefs about the
parameters of interest. Further work might consider different design
of observation times, leading to unbalanced longitudinal data, or wider seroconversion intervals. The seroconversion time for the new individual is given a uniform prior which reflects our belief that seroconversion is equally likely to occur at any time between the last negative and the first positive HIV test date. If information on testing behaviour is available, it may be incorporated in the choice of a different prior distribution.

Throughout this paper, we assume that all the information for the in-sample
individuals is available. In practice, we may not have access to all
data for the in-sample individuals but we may only know the posterior
distribution of the growth model parameters. In this case, a two-stage
analysis would be more applicable, where this
posterior distribution is used as a prior
distribution in the second stage.

Despite the further research required, we have provided a
  valuable proof-of-concept that fully Bayesian linear and non-linear
  mixed effects models for multiple biomarkers may be combined in joint models to improve
  estimation of the recency of HIV infection.
  
  \section*{Acknowledgements}

The authors would like to thank Dr Shaun Seaman, Dr Brian Tom, Dr Alex Welte, Dr Eduard Grebe, and the CEPHIA group for helpful discussions. This work was supported by the Medical Research Council [Unit Programme Number U105260566]; Public Health England; and the NIHR HPRU in Evaluation of Interventions.\vspace*{-8pt}


\end{document}